\documentclass[12pt]{article}

\pdfoutput=1

\usepackage[toc,page]{appendix}

\usepackage{color}
\usepackage{graphicx}
\usepackage{empheq}
\usepackage{hyperref}
\usepackage{epsf}
\usepackage{graphicx,epsfig}
\usepackage{float}
\usepackage{caption}
\usepackage{subcaption}

\usepackage{mathrsfs}
\usepackage{amsmath}
\usepackage{amssymb}
\usepackage{epsfig}
\usepackage{cite}
\usepackage{color,colordvi}
\newcommand{\be}{\begin{eqnarray}}
\newcommand{\ee}{\end{eqnarray}}
\newcommand{\bi}{\begin{itemize}}
\newcommand{\ei}{\end{itemize}}

\newcounter{hran}

\def\MSbar{\relax\ifmmode\overline{\rm MS}\else{$\overline{\rm MS}${ }}\fi}
\def\del{\partial}

\def\d{\rm d}

\def\d{{\rm d}}

\def\sc{\mathscr{I}}
\def\bz{{\overline{z}}}

\def\tg{\widetilde{g}}

\def\O{\Omega}

\begin{document}
\thispagestyle{empty}
\vspace{5mm}
\vspace{0.5cm}
\begin{center}

\def\thefootnote{\fnsymbol{footnote}}

{\Large \bf 
BMS in Cosmology\\
\vspace{0.25cm}	
}
\vspace{1.5cm}
{\large  A. Kehagias$^{a}$ and A. Riotto$^{b}$}
\\[0.5cm]

\vspace{.3cm}
{\normalsize {\it  $^{a}$ Physics Division, National Technical University of Athens, \\15780 Zografou Campus, Athens, Greece}}\\

\vspace{.3cm}
{\normalsize { \it $^{b}$ Department of Theoretical Physics and Center for Astroparticle Physics (CAP)\\ 24 quai E. Ansermet, CH-1211 Geneva 4, Switzerland}}\\

\vspace{.3cm}


\end{center}

\vspace{2cm}

\hrule \vspace{0.3cm}
{\small  \noindent \centerline{\textbf{Abstract}} \\[0.2cm]
\noindent 
Symmetries play an interesting   role  in cosmology. They are useful  
in characterizing    the  cosmological perturbations  generated during inflation and  lead to consistency relations involving the soft limit
of the statistical correlators of large-scale structure dark matter and galaxies overdensities.
On the other hand, in observational cosmology the carriers of the information about these large-scale  statistical distributions
are  light rays  traveling on null geodesics.
Motivated by this simple consideration, we study the structure of null infinity and the associated   BMS symmetry in a cosmological setting. 
 For decelerating Friedmann-Robertson-Walker  backgrounds, for which  future null infinity exists,  we find that the BMS transformations which leaves  the asymptotic metric invariant to leading order. Contrary to the asymptotic flat case, the BMS transformations in cosmology generate
Goldstone modes corresponding to  scalar, vector  and tensor degrees of freedom which may exist at  null infinity and perturb the asymptotic data.
Therefore,  BMS transformations generate physically inequivalent vacua as they populate the universe at null infinity with these physical degrees of freedom. We also discuss the gravitational memory effect when cosmological expansion is taken into account.
In this case,  there are extra contribution to the gravitational memory due to the tail of the retarded Green functions which are supported not only on the light-cone, but also in its interior. The gravitational memory effect  can be understood also from an asymptotic point of view as a transition among   cosmological BMS-related vacua.

\vspace{0.5cm} 
 \hrule
\vskip 1cm

\def\thefootnote{\arabic{footnote}}
\setcounter{footnote}{0}


\baselineskip= 15pt

\newpage 
\tableofcontents
\baselineskip=18pt

\section{Introduction}
Spacetimes with  asymptotic boundaries may possess diffeomorphisms that act non-trivial on boundary data. These ``large" diffeomorphisms are a subgroup of the full group of diffeomorphisms and they are referred to as  the asymptotic symmetry group.
The latter generates
asymptotic symmetries which are those diffeomorphisms of spacetime that transform asymptotically flat metrics again to asymptotically flat metrics. Strictly speaking, they are not symmetries of spacetime, that is transformations that leave the metric form invariant. Indeed, the metric is not invariant under such transformations, but the asymptotic symmetries  are nevertheless the closest to a symmetry in the asymptotic sense.  

The study of these symmetries started with 
the work by Arnowitt, Deser and Misner \cite{ADM}, who showed that energy, momentum and angular-momentum are associated with asymptotic symmetries at spatial infinity of an asymptotically Minskowski spacetime. Similarly, Bondi, Meissner and Sachs (BMS) \cite{bms1} considered  the asymptotic symmetries at future null infinity.  

The structure of null infinity is more subtle than that of spatial infinity since radiation reaches null infinity and distorts the asymptotics. 
BMS  found that these symmetries form in fact a much larger,  infinite dimensional  group now known as the BMS group \cite{bms1,bms2,Barut,Harvey,NU,bms3 }. 
The latter acts on null infinity and accounts for a large degeneracy of the asymptotic vacuum. The existence of such a group is counterintuitive as one expects that asymptotically, where spacetime is flat, the Poincar\'e group should be recovered. However, this is true for spacelike infinity, but not for null infinity (provided that  spacetime possesses one). 

The symmetry  group for null infinity 
 contains the Poincar\'e group as a subgroup,  but it also contains an infinite dimensional abelian subgroup parametrizing the so-called  supertranslations. The semidirect product of the supertranslations with the  Lorentz group form an infinite dimensional group, the BMS group. The (orthochronous) Lorentz group is in fact isomorphic to the SL(2,C)/${\mathbb Z}_2$ which generates global conformal (projective) transformations on the Riemann sphere. Relaxing global restrictions, the most general BMS transformation is a semidirect product of supertranslations and local conformal transformations on the sphere\cite{barnich}.

Interestingly,  being characterized by large  diffeomorphisms,  the BMS group   transforms asymptotically flat
solutions to new, physically inequivalent solutions.   As we mentioned, physically this is understandable since the BMS transformation creates a graviton out of the vacuum and this graviton can reach null infinity and distort the asymptotics. 

The  infinite-dimensional subgroup of supertranslations  generates arbitrary  translations of retarded time depending on the coordinates of the  sphere at null infinity. Asymptotically flat spacetimes possess two BSM symmetries acting on the future and past null infinity. One question connected with these symmetries is if they can be symmetries of the S-matrix in a scattering process. As  proven in a series of  interesting papers   \cite{strom1,strom1-1,strom2,strom3,strom4,strom5}, the diagonal BMS subgroup of the two BMS groups acting in past and future null infinity is indeed a symmetry of the S-matrix. Therefore, there exists a corresponding Ward identity which is nothing else that Weinberg's graviton soft theorem.   
The soft graviton is  just the  Goldstone boson associated to the breaking of supertranslations as the latter does not leave the vacuum  invariant.

BMS transformations are also connected to gravitational memory \cite{m1,m2,m3,m4}. It is known that  the passage of a gravitational wave burst  through a pair of nearby inertial detectors produces oscillations in their relative positions.  In other words,  a permanent  distortion is left behind a gravity wave train passing through a region of spacetime.  This has the effect that,  after the wave has passed, the detectors do not return to their initial position, but they acquire an extra displacement, known as gravitational memory. 
This effect can be understood also from an asymptotic point of view as a transition among   BMS-related vacua \cite{strom2,strom4}:   a supertranslation generates a shift in the spacetime metric, giving  rise to a displacement of nearby points. The BMS transformations may play a significant role also in the black hole information paradox \cite{ip1,ip2}.

Symmetries play also  crucial  role  in cosmological settings.
For instance, they are particularly useful   in  characterizing   the  cosmological perturbations  generated during inflation  \cite{lrreview} when   the de Sitter isometry group acts  as conformal group  on $\mathbb{R}^3$ for  the fluctuations  on super-Hubble scales. There the  SO(1,4) isometry
of the de Sitter background is realized as conformal symmetry of the flat $\mathbb{R}^3$ sections  and   correlators are constrained by conformal invariance \cite{antoniadis,creminelli1,us1,us2}. This  applies in the case in which the cosmological perturbations are generated 
by light scalar fields other than the inflaton (the field that drives inflation). In the opposite case of single-field inflation,   conformal consistency relations among the inflationary correlators
 have also been   investigated  \cite{creminelli2,hui,baumann1,baumann2}. The fluctuations in single-field inflation  are Goldstone bosons of a spontaneously broken dilation symmetry.  Being non-linearly realized, the broken symmetry is
still respected in Ward identities and leads to a relation between the variation of the $n$-point function of the comoving curvature perturbation $\zeta$ under dilation and  the squeezed limit of the $(n + 1)$-point function \cite{mal,cz}. These identities may be extremely useful in discriminating among the various mechanisms for the generation of the cosmological perturbations. For instance, the detection of a sizable primordial three-point correlator  in the squeezed limit would rule out all single-field  models where inflation is
driven by a single scalar field with canonical kinetic energy and an initial Bunch-Davies vacuum. 

When perturbations re-enter the horizon, they provide the seeds for the large-scale structure of the universe. The symmetries enjoyed by the Newtonian equations of motion of the non-relativistic dark matter   fluid coupled to gravity lead to   consistency relations  involving the soft limit of the $(n+1)$-correlator functions of dark matter   and galaxy overdensities \cite{cr1,cr2}.  They can be extended to the relativistic case and in single-field models of inflation since
in such a case the soft mode perturbations  can be seen as   diffeomorphisms \cite{cr3,cr4,cr5}. 
As for the  inflationary consistency relations, the ones for the large-scale structure  are a form of soft-pion theorem which relates
an $n$-point function to an $(n+1)$-point function where the additional leg represents the emission (or absorption) of the 
 Goldstone boson associated with a non-linearly realized symmetry. In particular, for large-scale structure the role of the Goldstone boson  is played by  the  peculiar
velocity which is  shifted in a non-linear way under some symmetry transformation, while
the same transformation shifts the density contrast  only linearly. The corresponding   consistency relations have the virtue of being true  for the galaxy overdensities,
independently of the bias between galaxy and dark matter.  

On the other hand, when dealing with observational cosmology, one should remember that information about the statistical distributions of the large-scale structure is obtained by collecting light rays emitted from the various objects, {\it e.g.} galaxies. Cosmologists (or better, their detectors) look at regions collecting light rays coming out from null, not spacelike, infinity. So, it is really at null infinity that one should
 investigate symmetries.

Based on these considerations, in  this paper we  discuss the BMS transformations in a cosmological setting. 
 In particular, we study  the structure of asymptotic past and future infinity in cosmological Friedmann-Robertson-Walker (FRW) spacetimes. For decelerating FRW backgrounds where future null infinity exists,  we find the corresponding BMS transformations. These are transformations that act at future null infinity and leave the asymptotic metric invariant to leading order. We also discuss the gravitational memory when cosmological expansion is taken into account. In this case,  there are extra contribution to the gravitational memory due to the tail of the retarded Green functions which are supported not only on the light-cone, but also in its interior.

Similar to what we have mentioned above, the BMS transformations in cosmology generate
Goldstone modes. Differently from the flat case though, they  correspond to not only tensor modes, but also to scalar and vector degrees of freedom.  They   may exist at infinity and perturb the asymptotic data. In this regard, BMS transformations generate physically inequivalent vacua as they populate the universe at null infinity with physical degrees of freedom. Furthermore, cosmological gravitational memory is connected to soft theorems. It would be interesting to investigate if  these considerations lead 
to new interesting consistency relations (soft theorems) for cosmological observables  \cite{kr}.

The paper is organized  as follows. In section 2 we present the structure of the 
asymptotic infinity for general FRW spacetimes. In section 3, we discuss the structure of null infinity for those FRW spacetimes that possess  one and their corresponding symmetries. In 
section 4, we present the BMS transformations in such cosmological backgrounds. In section 5 we discuss the gravitational memory in the presence of a  cosmological expansion, and
finally we conclude in section 6. Appendix A presents the Einstein equations near null infinity. Appendix B presents the retarded Green function in FRW space times necessary to discuss the gravitational memory effect. Appendix C introduces null tetrads to make use of the Newman-Penrose (NP) formalism needed to discuss the gravitational memory.  

\section{Structure of Infinity}
 
We will review below the structure of asymptotic infinity of cosmological backgrounds by using conformal diagrams. We will see that inflating universes do not possess a null infinity and therefore radiation cannot be detected, as expected, at infinity. On the contrary, spacetimes representing decelerated expanding universes do have null infinity as radiation can reach infinity. In order to set up the notation, we start with the simplest static Minkowski space-time.  The expert reader can skip this part.

\subsection{Minkowski spacetime}     
Let us consider four-dimensional Minkowski spacetime with metric in polar coordinates
\begin{eqnarray}
\d s^2=-\d t^2+\d r^2+r^2\, \d \Omega_2^2, \label{MM}
\end{eqnarray}
where $\d \Omega_2^2=\d\theta^2+\sin\theta^2\, \d\phi^2$ is the metric on $\mathbb{S}^2$
and 
\begin{eqnarray}
-\infty< t<\infty , ~~~0\leq r<\infty.
\end{eqnarray}
We can also write the flat metric Eq. (\ref{MM}) in Bondi coordinates $(u=t-r, r,\theta,\phi)$ and $(v=t+r,r,\theta,\phi)$
 as 
 \begin{eqnarray}
 \d s^2=-\d u^2-2\d u \d r +r^2\, \d \Omega_2^2=-\d v^2+2\d v \d r +r^2\, \d \Omega_2^2. 
 \end{eqnarray}
By defining new coordinates 
\begin{eqnarray}
&&U=\arctan(t-r)=\arctan(u)=\arctan(v-2r), \nonumber \\
&&V=\arctan(t+r)=\arctan(v)=\arctan(u+2r),
\end{eqnarray}
with 
\begin{eqnarray}
-\pi/2 < U<\pi/2, ~~~-\pi/2 < V<\pi/2, ~~~U\leq V,
\end{eqnarray}
we can bring infinity at a finite distance. Indeed, we may express the flat Minkowski 
metric as 
\begin{eqnarray}
\d s^2=\frac{1}{4 \cos^2 V\cos^2 U}\left(-\d T^2+\d R^2+\sin^2R\,\d \Omega_2^2\right),
\end{eqnarray}
where 
\begin{eqnarray}
T=U+V, ~~~R=V-U
\end{eqnarray}
and
\begin{eqnarray}
0\leq R<\pi, ~~~ |T|+R<\pi. 
\end{eqnarray}
Therefore, the original Minkowski metric is  conformally related to the metric  $\d\widetilde s^2$ of the Einstein static universe as
\begin{eqnarray}
\d s^2=\Omega^2 \d\widetilde s^2, 
\end{eqnarray}
with $\Omega^{-1}=2 \cos V\cos U$ and 
\begin{eqnarray}
\d\widetilde s^2=-\d T^2+\d R^2+\sin^2R\, \d \Omega^2.  \label{eu}
\end{eqnarray}
The sections $T=$ constant are three-dimensional spheres $\mathbb{S}^3$ so that the topology is $\mathbb{R}\times \mathbb{S}^3$. 
The conformal diagram of Minkowski spacetime is drawn in Fig.1, where we have indicated:
\begin{eqnarray}
i^+&=&\mbox{ future timelike infinity}~ ~ (T=\pi , ~R=0),~~~~(u=\infty,v=\infty,~r=0)\nonumber \\
i^-&=&\mbox{ past timelike infinity}~ ~  ~ ~ \,(T=-\pi , ~R=0),~~~(u=-\infty,v=-\infty,~r=0)\nonumber \\
i^0&=&\mbox{ spatial infinity}~ ~  ~ ~ ~  ~ ~ ~ ~ ~ ~ ~ ~(T=0 , ~R=\pi),~~~(u=\infty, v=\infty, ~r=\infty)\nonumber \\
\sc^+&=&\mbox{ future null infinity}~ ~ ~ ~ ~ ~ ~ ~ (T=\pi-R , ~0<R<\pi), ~~~(-\infty<u<\infty,~r=\infty)\nonumber \\
\sc^-&=&\mbox{ past null infinity}~ ~ ~ ~ ~ ~ ~ ~  ~ ~ ~(T=-\pi+R , ~0<R<\pi),~~~ (-\infty<v<\infty,~r=\infty).\nonumber\\
&&
\end{eqnarray}
Note that since $R=0$ and $R=\pi$ are just the north and south poles of the $\mathbb{S}^3$, $i^+,i^-,i^0$ are actually points. On the other hand, $\sc^+,\sc^-$ are null surfaces corresponding to future and past null infinity. All null geodesics start at $\sc^-$ and  end at $\sc^+$, all timelike geodesics start at $i^-$ and end at $i^+$, whereas, spacelike ones start at $i^0$ and end at $i^0$.

\subsection{FRW spacetime}

Let us now consider the spatially flat FRW universe dominated by a fluid with equation of state $w$. The corresponding geometry has the  line element 
\begin{eqnarray}
\d s^2=-\d t^2+a^2(t)\bigg(\d r^2+r^2\,\d\O_2^2\bigg),\,\,\,\,\,\,\,a(t)=\left(\frac{t}{t_0}\right)^{\frac{2}{3(w+1)}}. \label{dsfrw}
\end{eqnarray}
In conformal time $\d\tau=\d t/a(t)$, we may write (\ref{dsfrw}) as
\begin{eqnarray}
\d s^2=\left(\frac{\tau}{\tau_0}\right)^{2q}\bigg(-\d \tau^2+\d r^2+r^2d\O_2^2\bigg)=\left(\frac{u+r}{L}\right)^{2q}\bigg(-\d u^2-2\d u\d r+r^2\, \d\O_2^2\bigg),
\end{eqnarray}
where the Bondi coordinates are now $(u=\tau-r,r,\theta,\phi)$ and  $q=2/(3w+1)$.  Now, there exists two distinct cases\footnote{As known, our  universe has gone through various phases with different equation of states. Here we deal with the case of constant $q$ for simplicity, but one will have to cope with the fact that the universe has experienced a transition  from a decelerating to an accelerating phase at a redshift of order unity when investigating the possible role of BMS transformations.}:
\vskip 0.5cm
\noindent
\centerline{{\bf Deceleration}}
\vskip 0.3cm
\noindent
In this case, we have  
\begin{eqnarray}
-\frac{1}{3}<w\leq 1,\,\,\,\, q>0.
\end{eqnarray}
The range of $\tau$ is $\tau\in (0,\infty)$. At $\tau=0$ we have a spacelike singularity and future infinity is at $\tau=\infty$. 
We may express $(\tau,r)$ in terms of   new coordinates $(T,R)$  as
\begin{eqnarray}
\tau=\frac{\sin T}{\cos R+\cos T}, ~~~r=\frac{\sin R}{\cos R+\cos T},
\end{eqnarray}
where 
\begin{eqnarray}
0<T<\pi, ~~~0<R<\pi-T.
\end{eqnarray}
In these coordinates 
we may write (\ref{dseta}) as
\begin{eqnarray}
\d s^2=
  \frac{\sin^{2q} |T|}{4\left(\cos \frac{R+T}{2}\cos\frac{R-T}{2}\right)^{2+2q}}\bigg(-\d T^2+\d R^2+\sin^2R\, d\O_2^2\bigg),\label{dseta}
\end{eqnarray}
which is conformal to metric in Eq.  (\ref{eu}). There exists a Big-Bang singularity at past (spacelike)  infinity $\sc^-$, a future timelike infinity at $(T=\pi,R=0)$ and a future null infinity $\sc^+$ at $T+R=\pi$ ($-\infty<u<\infty$, $r=\infty$) as dictated on the right of Fig. 1. 
\begin{figure}[h!]
\centering
\begin{subfigure}{.5\textwidth}
  \centering
  \includegraphics[width=.6\linewidth]{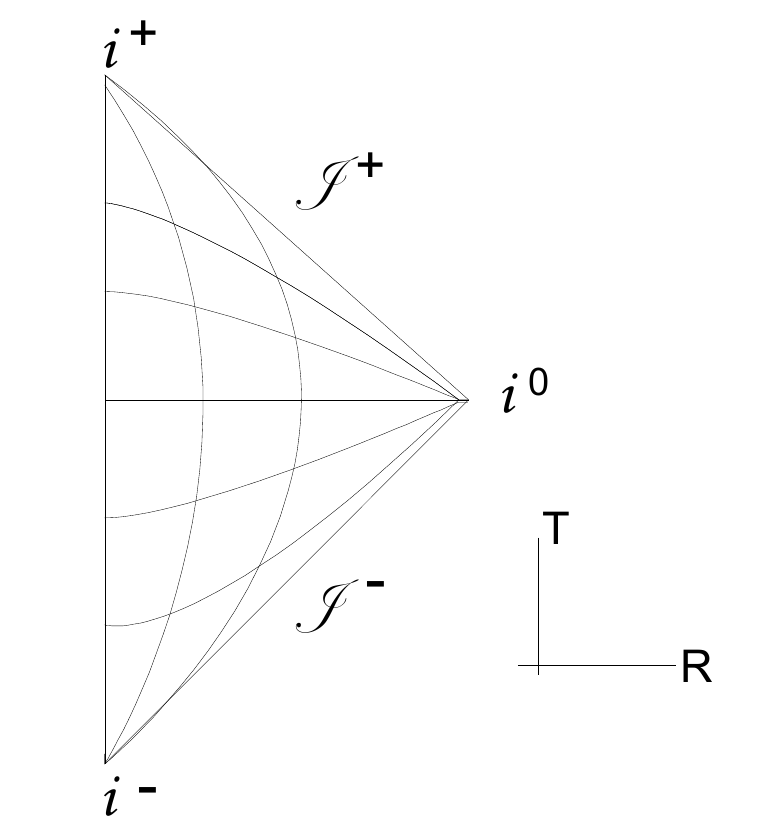}
  \label{fig:sub1}
\end{subfigure}%
\begin{subfigure}{.5\textwidth}
  \centering
  \includegraphics[width=.6\linewidth]{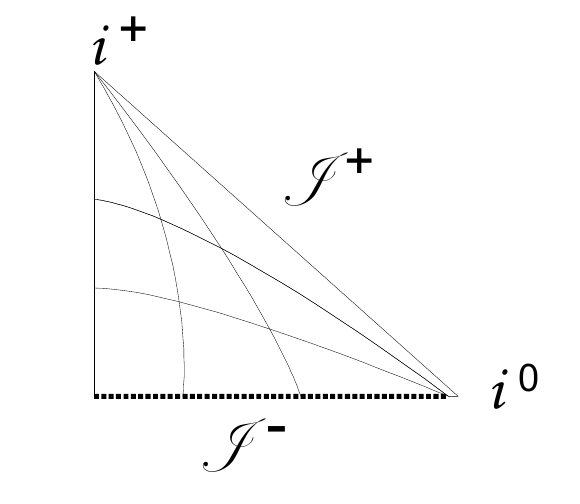}
  \label{fig:sub2}
\end{subfigure}
\caption{\small Conformal diagrams for Minkowski (left) and decelerating FRW (right) spacetimes.}
\label{fig:test}
\end{figure}

\vskip 0.5cm
\noindent
\centerline{{\bf Acceleration}}
\vskip 0.3cm
\noindent
In this case, we have 
\begin{eqnarray}
-1<w<-\frac{1}{3},\,\,\,\, q<0,
\end{eqnarray}
and the range of $\tau$ is now $\tau\in (-\infty,0)$. At $\tau=-\infty$ we have a spacelike singularity and future infinity is at $\tau=0$. Similarly, the range of  the coordinates $(T,R)$ defined above is now
\begin{eqnarray}
-\pi<T<0,~~~0<R<\pi+T.
\end{eqnarray}
The past infinity $\sc^-$ is null  and singular, whereas the future infinity $\sc^+$ is spacelike.
The Penrose diagram is drawn in Fig.2. 

 \vskip 0.5cm
\noindent
\centerline{{\bf Limiting cases}}
\vskip 0.3cm
\noindent
There are two limiting cases described by $w=-1/3$ and $w=-1$, the latter being  de Sitter spacetime. The corresponding diagrams are given in Fig. 3. Only half of   de Sitter has been indicated, the causal part of an observer at the left hand side (the north pole). Note that the future infinity $\sc^+$ is spacelike for de Sitter, whereas  $\sc^-$ is the particle horizon and it is null.  
Finally, the Penrose diagram for the $w=-1/3$ case is similar to the Minkowski one in Fig.1, where  the past null infinity $\sc^-$ is singular now.

\begin{center}
\begin{figure}
\begin{center}
\includegraphics[scale=.8]{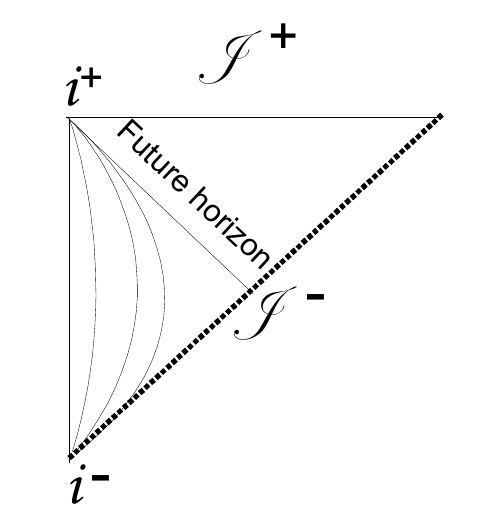}
\caption{\small Conformal diagram for accelerating FRW cosmology.}
\end{center}
\end{figure}
\end{center}
\begin{figure}[H]
\centering
\begin{subfigure}{.5\textwidth}
  \centering
  \includegraphics[width=.6\linewidth]{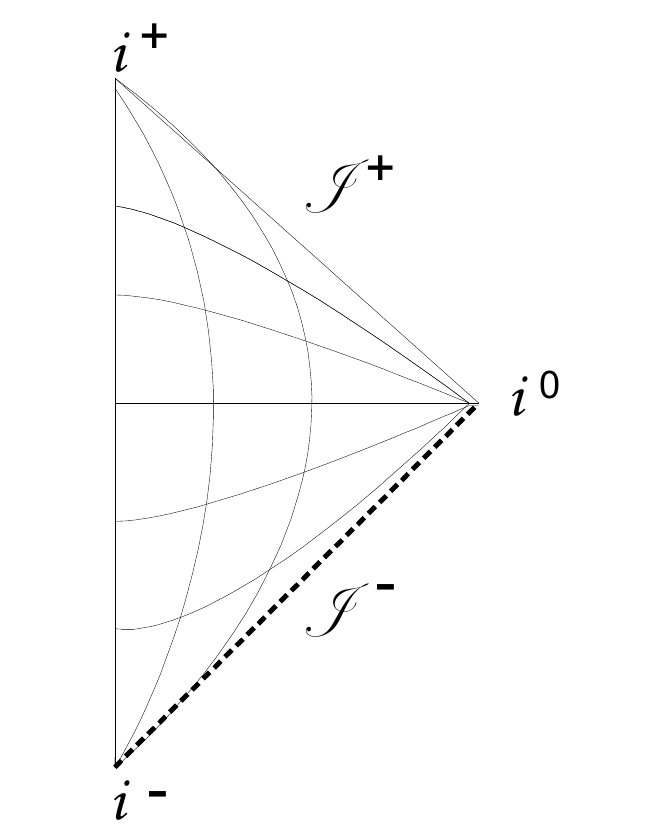}
  \label{fig:sub1}
\end{subfigure}%
\begin{subfigure}{.5\textwidth}
  \centering
  \includegraphics[width=.6\linewidth]{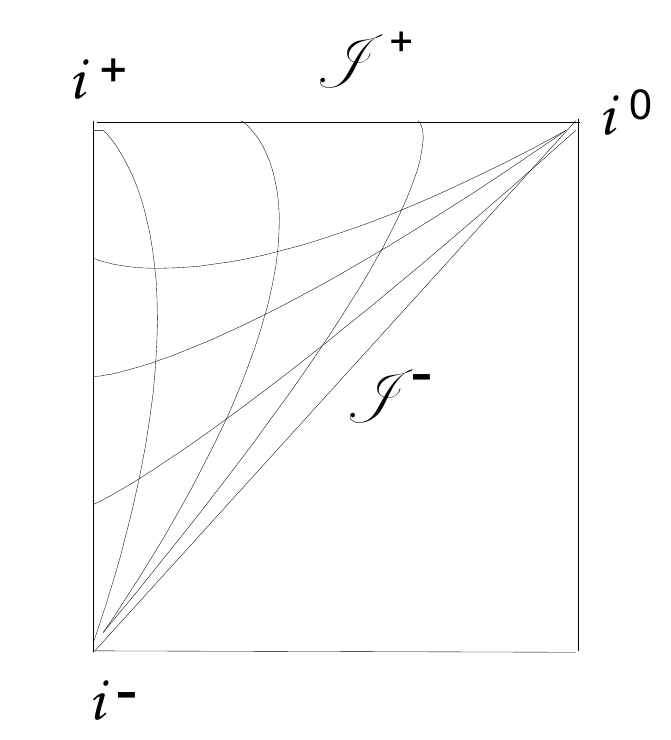}
  \label{fig:sub2}
\end{subfigure}
\caption{\small Conformal diagrams for FRW geometry with $w=-1/3$ (left) and (the causal half part of the) de Sitter  spacetime (right).}
\label{fig:test}
\end{figure}



\section{Null Infinity}
In this section,  and for the benefit of the reader, we summarize the main features of asymptotic space times at null infinity. 

Let  $({\cal M}, g_{\mu\nu})$ be   a four-dimensional spacetime. The manifold  $\widetilde{\cal M}$ with  boundary $\sc$ and metric $\tg_{\mu\nu}$    is  asymptotic to ${\cal M}$ if ${\cal M}$ is diffeomorphic to $\widetilde{\cal M}-\sc$ and  the following  two conditions are satisfied:
there exists  a smooth function $\O$ on $\widetilde{\cal M}$ such that 
\begin{eqnarray}
&&{\rm i)}~~~\tg_{\mu\nu}=\Omega^2 g_{\mu\nu}   ~~\mbox{on}\,\,\, \widetilde{\cal M},\\
&&  {\rm ii)}~~
\O=0,~~~\nabla_\mu \O\neq 0 ~~\mbox{and}~~~\tg^{\mu\nu}\nabla_\mu \O\nabla_\nu \O=0 ~~~
\mbox{on}\,\,\, \sc. 
\end{eqnarray}
Let $\xi^\mu$ now be a vector field in ${\cal M}$, which is not a Killing vector, {\it i.e.} ${\cal L}_\xi g_{\mu\nu}\neq 0$. Then, $\xi^\mu$ will generate asymptotic symmetries  if
\begin{eqnarray}
\Omega^2 {\cal L}_\xi g_{\mu\nu}=0,  \label{lie}
\end{eqnarray}
which is the closest  to a symmetry at  infinity. Using well know identities, we may express 
(\ref{lie}) as
\begin{eqnarray}
0=\Omega^2 {\cal L}_\xi g_{\mu\nu}= {\cal L}_\xi \widetilde{g}_{\mu\nu}-2\Omega^{-1} \xi^\mu 
\nabla_\mu \Omega \widetilde g _{\mu\nu}.
\end{eqnarray}
Therefore, we find that 
\begin{eqnarray}
{\cal L}_\xi \widetilde{g}_{\mu\nu}=\psi  \widetilde g _{\mu\nu}, ~~~\psi=2\Omega^{-1} \xi^\mu  \label{k1}
\nabla_\mu \Omega.
\end{eqnarray}
If the  vector field $\xi^\mu$ does not vanish at $\sc$, it would generate conformal diffeomorphisms at $\sc$ and it would be a conformal Killing vector. In other words, asymptotic symmetric at $\sc$ are generated by conformal Killing vectors. We can easily also find the condition for the non-vanishing of $\xi^\mu$. From (\ref{k1}), we see that we should demand that
$\psi$ is a smooth function and therefore
  we must have 
\begin{eqnarray}
2\xi^\mu \nabla_\mu \Omega= \Omega \psi . 
\end{eqnarray}
We will  consider now  spacetimes with null future infinity $\sc^+$ such as  Minkowski or  FRW backgrounds  with $-1/3<w\leq 1$.  The metric for  Minkowski space-time can be written  in terms of Bondi coordinates $(u=t-r,r,\theta,\phi)$ as
\begin{eqnarray}
\d s^2=-\d u^2-2 \d u \d r+r^2\, \d\O_2^2.  \label{r}
\end{eqnarray}
The vector $n=n^\mu \del/\del x^\mu=\partial/\partial r $
is normal to the surface $u=$ constant  and it  is 
 null $(g_{\mu\nu}n^an^a=0)$ so that the surfaces $u=$ constant  are  null. Null infinity is therefore at $r=\infty$ which cannot be captured by the metric (\ref{r}). Therefore, we may consider the unphysical metric 
\begin{eqnarray}
 \d\widetilde s^2=\frac{1}{r^2}\d s^2=-\frac{\d u^2+2 \d u\ d r}{r^2}+\d\O_2^2,
 \end{eqnarray} 
 in which the $r=$ constant  sections  will have the induced three-dimensional metric 
 \begin{eqnarray}
  d\widetilde s_3^2=- \left(\frac{\d u^2}{r^2}\right)+\d \Omega_2^2.
  \end{eqnarray} 
In particular, at $r=\infty$ we will have 
\begin{eqnarray}
\d\widetilde s_3^2=\d\theta^2+\sin^2\theta\, \d\phi^2+0\cdot \d u^2. \label{mj}
\end{eqnarray}
We would like now to find all transformations that preserve the conformal metric on $\sc^+$. These should include the conformal maps of the 
$\mathbb{S}^2$ to itself and, in particular,  the transformations
\begin{eqnarray}
&&\theta\to \theta,\\
&&\phi\to \phi,\\
&&u\to F(u,\theta,\phi).
\end{eqnarray}
Indeed, after covering $\mathbb{S}^2$ by a complex coordinate $z$ defined as
\begin{eqnarray}
z=\cot \frac{\theta}{2}e^{i\phi},
\end{eqnarray}
the metric (\ref{mj}) is written as 
\begin{eqnarray}
\d\widetilde s_3^2=4\frac{\d z\ d\overline z}{(1+z\overline z)^2}+0\cdot \d u^2.
\end{eqnarray}
As a consequence,  the transformations that preserve the conformal form of the metric (\ref{mj}) is the Newman-Unti transformation \cite{NU}
\begin{eqnarray}
&&z\to \frac{az+b}{cd+d}.\\
&&u\to F(u,z,\overline z). \label{NU}
\end{eqnarray}
Note that we have assumed that the transformation does not change the orientation of $u$,  (which will interchange $\sc^+$ with $\sc^-$) so that $\partial F/\partial u>0$. Allowing for spatial reflections on $\mathbb{S}^2$, which is also a conformal mapping, the transformation $z\to\overline z$ should be included.  

The Newman-Unti transformations are the most general transformations allowed. We may restrict it by imposing appropriate  conditions. One  condition is the following. Let us recall that conformal transformations preserve the angle between vectors. One may also define a ``null angle" as the ratio $\d u/\d\O_2$ \cite{Barut}.  We may then search for transformations of $\sc^+$ that also preserve the null angle. In other words, we look for transformations  for which the  rescalings of $\d\O_2$ are compensated by corresponding  rescalings of $u$. These transformations should be of the form (\ref{NU}), but with the function $F$ such that  the ratio $\d u/\d\O_2$ is left invariant.
Since under conformal transformations we have for $\d\O_2$
\begin{eqnarray}
\d \Omega_2\to K \d \Omega_2,
\end{eqnarray}
where $K=K(z,\overline z)>0$, we should also demand 
\begin{eqnarray}
\d u \to K u, 
\end{eqnarray}
which can be integrated to give
\begin{eqnarray}
u\to K\Big{(} u+ \alpha(z,\overline z)\Big{)}.
\end{eqnarray}
Therefore, the symmetries of the null infinity $\sc^+$ comprises 
\begin{eqnarray}
&&z\to \frac{az+b}{cd+d}\label{BMS1}\\
&&u\to K\Big{(} u+ \alpha(z,\overline z)\Big{)}. \label{BMS2}
\end{eqnarray}
The transformations (\ref{BMS1}) and (\ref{BMS2}) form the so-called BMS group $\mathscr{G}$.
The transformation (\ref{BMS1}) forms the conformal group in two-dimensions, which is isomoprhic to the orthochonous Lorentz group O(1,3). In other words, the Lorentz group acts on the two-dimensional sphere by a conformal transformation  like (\ref{BMS1}). 
In particular, the transformation 
\begin{eqnarray}
&&z\to z \label{st1},\\
&&u\to \Big{(} u+ \alpha(z,\overline z)\Big{)}, \label{st}
\end{eqnarray}
 that does not involve a Lorentz rotation, is called supertranslation. 
We may expand $a(z,\overline z)$ in spherical harmonics as 
\begin{eqnarray}
\alpha=\sum_{m,\ell}\alpha_{\ell m}Y_{\ell m}(\theta,\phi).
\end{eqnarray}
Transformations that involve only the lowest $\ell=0,1$ components are written as
\begin{eqnarray}
\alpha=\alpha_0 +\alpha_1 \sin\theta\cos\phi+\alpha_2 \sin\theta\sin\phi+\alpha_3\cos\theta,
\end{eqnarray}
which, in terms of $z,\overline z$ is written as 
\begin{eqnarray}
\alpha=\frac{b_0+b_1 z+\overline b _1 \overline z+b_2 z \overline z}{1+z \overline z}.
\end{eqnarray}
This is a four-parameter transformation, a translation, and it is actually the transformation induced on $\sc^+$ by standard translations of the Minkowski spacetime. 
Supertranslation form a normal subgroup $\mathscr{S}$ of the  the BMS group $\mathscr{G}$
and  the quotient group $\mathscr{G}/\mathscr{S}$ is the conformal group on the two-sphere $\mathbb{S}^2$ as can be seen from Eqs.  (\ref{BMS1}) and (\ref{BMS2}). 

Note that the above discussion is  valid for  
both $\sc^+$ and $\sc^-$. The  topology of $\sc^\pm$ is  of the form $\mathbb{S}^2\times \mathbb{R}$ where $\mathbb{R}$ is formed by the null generators of $\sc^\pm $ . As a result, $\mathbb{S}^2$ sections of $\sc^+$ or $\sc^-$ 
are related by a conformal transformation of the $\mathbb{S}^2$. Without loss of generality, we may assume that the conformal factor $\Omega$ has been chosen such that the unphysical metric $\d\widetilde s^2$ is the standard metric on the unit round $\mathbb{S}^2$. 
This is possible since for any $\Omega$ we can define a new $\Omega'=\omega\cdot \O$ 
where $\omega>0$ is a smooth function on $\sc^\pm$. The new $\O'$ will have the required properties and will allow to rescale the metric on $\sc^\pm$ at will.  Thus, without loss of generality we may assume that the unsphysical metric  
has the form (\ref{mj}).

\subsection{Null infinity for FRW cosmologies}
Let us now consider a FRW geometry for a fluid with $-1/3<w\leq 1$. In this case the metric, as we already wrote,  is expressed in  $(u,r,\theta,\phi)$  coordinates as  ($q>0$)
\begin{eqnarray}
\d s^2=\left(\frac{u+r}{L}\right)^{2q}\bigg(-\d u^2-2\d u\d r+r^2\, \d\O_2^2\bigg).\label{dseta1}
\end{eqnarray}
Note that the usual cosmological parameters are  expressed now  through these  coordinates. For example, the Hubble parameter ${\cal H}$ and the redshift $z$ will be 
\begin{eqnarray}
{\cal H}=\frac{\del_\tau a}{a}=\frac{\del_u a}{a}=\frac{q}{u+r}, ~~~1+z=\frac{(u+r)^q|_{\cal O}}{(u+r)^q|_{\cal E}},
\end{eqnarray}
where ${\cal O}$ and ${\cal E}$ indicate the observer and the emitter location, respectively.
The sections $u=$ constant are null and the future null infinity $\sc^+$ is reached at $r=\infty$. Using the above arguments, we may take the unphysical metric to be 
\begin{eqnarray}
\d\widetilde s^2=\Omega^2 \d s^2\, , ~~~~\Omega^2=
\left((u+r)/\tau_0\right)^{-2q}r^{-2},
\end{eqnarray}
so that the metric on $\sc^+$ is written as 
\begin{eqnarray}
d\widetilde s_3^2=4\frac{dzd\overline z}{(1+z\overline z)^2}+0\cdot \d u^2. \label{fr}
\end{eqnarray}
The symmetries of null infinity are then also given by the Newman-Unti group (\ref{NU}) and the BMS transformations (\ref{BMS1}) and (\ref{BMS}) for the null-angle preserving ones. 

Note that for FRW background with $-1\leq w<-1/3$,  ($q<0$), the metric is written as 
\begin{eqnarray}
\d s^2=\left(-\frac{L}{u+r}\right)^{2|q|}\bigg(-\d u^2-2\ du\d r+r^2\, \d\O_2^2\bigg).\label{dseta2}
\end{eqnarray} 
Again, the sections $u=$ constant are null and future infinity $\sc^+$ is at $r+u=0$, which is spacelike.

\section{BMS Transformations in FRW spacetimes}

We will now proceed to find the asymptotic symmetries of FRW backgrounds. In particular, we are interested in extending the asymptotic symmetries from $\sc^+$ to the interior of spacetime.  For this, let us first write  the FRW metric  (\ref{dseta1}) 
in Bondi coordinates  $(u,r,z,\bz)$ as 
($q>0$) 
\begin{eqnarray}
\d s^2=a^2(u,r)\bigg(-\d u^2-2\,\d u \d r+2\,r^2\gamma_{z\overline z}\d z \d\overline{z}\bigg),\label{dseta11}
\end{eqnarray}
where 
\begin{eqnarray}
\gamma_{z\overline z}=\frac{2}{(1+z\bz)^2}
\end{eqnarray} 
is the metric on the unit sphere and 
\begin{eqnarray}
a(u,r)=\left(\frac{r+u}{L}\right)^q
\end{eqnarray}
is the scale factor written in Bondi coordinates. From now on we will deal with  the case $q>0$, which corresponds to a decelerating expansion, since in this case there is null infinity. In the opposite case ($q<0$), conformal infinity is spacelike and there is no radiation to be detected. 
Perturbations around the FRW metric (\ref{dseta11}) in Bondi coordinates  will be of the general form
\begin{eqnarray}
\d s^2=a^2\bigg(\overline{g}_{\mu\nu}+\delta g_{\mu\nu} \d x_B^\mu
\d x_B^\nu\bigg)=
a^2\bigg(-\d u^2-2\d u \d r+2r^2\gamma_{z\bz}\d z\ d\bz+\delta g_{\mu\nu} \d x_B^\mu
\d x_B^\nu\bigg), \label{mp}
\end{eqnarray}
where  $\delta g_{\mu\nu}=\delta g_{\mu\nu}(u,r,z,\bz)$. 
Under a diffeomorphism generated by $\xi^\mu=(\xi^u,\xi^r,\xi^z,\xi^\bz)$, the metric perturbations 
$\delta g_{\mu\nu}$ transforms as 
\begin{eqnarray}
\delta g_{\mu\nu}\to\delta   \widetilde g_{\mu\nu}=\delta g_{\mu\nu}+\partial_\rho \overline{g}_{\mu\nu}\xi^\rho+\overline{g}_{\mu\rho}\partial_\nu \xi^\rho+\overline{g}_{\rho\nu}\partial_\mu \xi^\rho,
\end{eqnarray}
where $\overline{g}_{\mu\nu}$ is the background FRW metric. In particular,  for the perturbations (\ref{mp}) we find 
\begin{eqnarray}
\delta\widetilde  g_{uu}&=& \delta g_{uu}-\frac{2q}{r}\xi^r-2 \xi^r,_u-2 \xi^u,_r,
\\
\delta \widetilde g_{ur}&=& \delta g_{ur}-\frac{2q}{r}\xi^r-2\xi^r,_r-2\xi^u,_r-2 \xi^u,_u\\
\delta \widetilde g_{uz}&=&\delta g_{uz}-\xi^r,_z-\xi^u,_z+r^2 \xi_{z},_u, \\
\delta \widetilde g_{rr}&=&\delta g_{rr}+2 \xi^r,_r,\\
\delta \widetilde g_{rz}&=& \delta g_{rz}-\xi^u+r^2 \xi_z,_r,\\
\delta \widetilde g_{zz}&=&\delta g_{zz}+2D_z\xi_z,\\
\delta \widetilde g_{z\bz}&=&\delta g_{z\bz }+r\bigg[2(1+q) \xi^r\gamma_{z\bz}+r(D_z\xi_\bz+D_\bz \xi_z)  \bigg],
\end{eqnarray}
where  $\xi_{z}=\gamma_{z\bz}\xi^\bz$.
Then by selecting the vector $\xi^\mu$ as 
\begin{eqnarray}
&&\xi^u=\delta g_{rz}-r^2 \xi_z,_r, \\
&&\xi_r=\frac{r}{2q}\left(\delta g_{ur}-\delta g_{rr}-2 \xi^u,_r-\xi^u,_u\right),\nonumber \\
&&\xi^r,_r=\frac{1}{2}\delta g_{rr},\label{kil}
\end{eqnarray}
we may choose a gauge 
similar to the Bondi gauge
\begin{eqnarray}
\delta g_{rr}=\delta g_{rz}=\delta g_{r\bz}=0. \label{Bg}
\end{eqnarray}
In the vicinity of null infinity $\sc^+$ at $r\to \infty$, we assume that the  metric perturbations can be expanded as 
\begin{eqnarray}
\delta g_{\mu\nu}(u,r,z,\bz)=\delta g_{\mu\nu}^{(0)}(u,z,\bz)+\frac{\delta g_{\mu\nu}^{(1)}(u,z,\bz)}{r}+{\cal O}(r^{-2}).
\end{eqnarray}
Then, in the   gauge  (\ref{Bg}),  the perturbed metric  can be expressed  as\footnote{The corresponding Einstein equations are found in Appendix A.}
\begin{eqnarray}
\d s^2&=&\left(\frac{r+u}{L}\right)^{2q}\bigg(-\d u^2-2\,\d u \d r+2\,r^2\gamma_{z\bz}\d z \d\bz +N\,\d u^2+\frac{2m}{r}\,\d u^2 +2\,r^2\gamma_{z\bz} C \d z  \d\bz\nonumber \\
&+&2E\, \d u \d r+\frac{2F}{r}\, \d u \d r+r\, C_{zz}\d z^2+r \,C_{\bz\bz}\d\bz^2+2r\,\gamma_{z\bz} C_{z\bz}\,\d z\d\bz-2\, U_z\d u \d z-2\, U_{\bz} \,\d u \d\bz+ \cdots
\bigg),\nonumber\\
&&
\label{dseta1111}
\end{eqnarray}
where $N,m,C,C_{zz},C_{\bz\bz}, C_{z\bz}, E,F$ are functions of $(u,z,\bz)$ only. Notice that all of them are scalars apart from 
$C_{zz}$ and $C_{\bz\bz}$, which parametrize the tensor mode and $B_i=(E+F/r)x_i/r$ which is a vector mode.
We will also use the notation 

\be
N_{zz}=\del_u C_{zz}
\ee
 for the Bondi news.   Note  that, dealing with an expanding universe and differently from the static case, we  had  included the zero mode perturbations $N(u,z,\bz)$ and $C(u,z,\bz)$ since  there can be scalar perturbations that do no  die off at null infinity. 

The transformations of the metric  (\ref{dseta1111}) that preserve its asymptotic form are given  by the infinitesimal BMS transformations (supertranslations)  \cite{bms1}
\begin{eqnarray}
&&u\to u-f, ~~~r\to r-D^zD_zf,\nonumber \\
&&z\to z+\frac{1}{r} D^z f, ~~~
\bz\to\bz+\frac{1}{r} D^\bz f, \label{BMS}
\end{eqnarray}
where 
\be
f=f(z,\bz),
\ee
 provided that 
 
 \be
 C(u,z,\bz)=0.
 \ee
 This choice is dictated by the fact that the transformation  (\ref{BMS}) generates an order one  $\delta g_{rz}$ term  
\begin{eqnarray}
\delta g_{rz}=-C D_z f. 
\end{eqnarray}
In order to preserve the  gauge (\ref{Bg}), this term has to vanish and this  leads to 
 $C=0$. 
The transformation in Eq. (\ref{BMS}) is generated by the vector field 
\begin{eqnarray}
{\bf \xi}= -f\del_u-D^z D_z f \del_r+\frac{1}{r}D^zf \del_z +\frac{1}{r}D^\bz f \del_\bz, \label{gBMS}
\end{eqnarray}
and leaves the asymptotic form of the metric invariant  by inducing a corresponding change in the asymptotic data as 
\begin{eqnarray}
\boxed{
\begin{array}{rcl}
\delta N&=&-f\del_u N,\label{sy}\\
 \delta m&=&-f \del_u m+\frac{1}{2} \left(D^zfD_z N+D^\bz f D_\bz N\right) \nonumber \\
 &-&  q N D^2 f -2 q f N +q D^2f, \label{lie1}\\
 \delta E &=&-f\partial_u E,\\
 \delta F&=& -f\partial_u F+D^zED_z f+D^\bz ED_\bz  f
 +2q(1-E)\left(D^2f+f\right),\\
 \delta C_{zz}&=&-f \del_u C_{zz}
 +2  D_z^2 f,\label{lie2}\\
\delta C_{z\bz}&=&-f \del_u C_{z\bz}
-qD^2f -2 q f, \label{lie3}\\
\delta U_z&=&-f \del_u U_z+ND_z f-D_z f- D_z D^2 f. 
 \label{sx}
 \end{array}
 }
 \end{eqnarray} 
From the equations above we can now read off the Goldstone modes of the BMS transformations in the cosmological setting. The quantity
$C_{zz}$ transforms non-linearly and it represents the usual Goldstone boson associated to the graviton mode; the quantities
 $m$ and $C_{z\bz}$ also transform non-linearly  and represent the scalar Goldstone modes. Their non-linear  terms appear only  for $q\neq 0$. Therefore,  $m$ and $C_{z\bz}$ can be considered as Goldstone bosons  only when a cosmological setting is considered. The same consideration applies to the vector degree of freedom generated by $F$
 
$$
\boxed{
\begin{array}{rcl}
C_{zz}&=&{\rm tensor}\,\,{\rm Goldstone}\,\,{\rm mode} \\
m\,{\rm and}\,C_{z\bz}&=&{\rm scalar}\,\,{\rm Goldstone}\,\,{\rm modes}\,\,{\rm only}\,\, \,\,{\rm for} \,\,q\neq 0\\
F&\Rightarrow&{\rm vector}\,\,{\rm Goldstone}\,\,{\rm mode}\,{\rm only}\,\, \,\,{\rm for} \,\,q\neq 0. \\
\end{array}
}
$$
Of course, one has to check if these Goldstone modes are physical (usually dubbed in cosmology adiabatic modes), that is if the corresponding function $f$ which generates them satisfies
Einstein's equations.

For instance, if we take the radiation-dominated case, $q=1$, and a constant $f$, one can easily show from Eqs. (\ref{sy})-(\ref{sx}) one generates from the vacuum the two  scalars mode and the vector mode

\be
C_{z\bz}=-m=-F=-2\,f,
\ee
and the corresponding metric

\be
\d s^2=a^2(u,r)\bigg[-\d u^2+\frac{4}{r}f\, \d u^2-2\,\d u \d r+\frac{4}{r}f\,\d u \d r+
2\,r^2\gamma_{z\overline z}\left(1-\frac{f}{r}\right)\d z \d\overline{z}\bigg].
\ee
One can easily check from Appendix A that this metric satisfies all Einstein's equations. 

\section{Cosmological gravitational memory}
The passage of a finite pulse of radiation or other forms of energy through a region of
spacetime produces a gravitational field which moves nearby detectors.   This effect is known as gravitational memory and its direct measurement  may be possible in the coming
decades \cite{favata}. The final positions
of a pair of nearby detectors are generically displaced relative to the initial ones according
to a simple and universal formula \cite{m1,m2,m3,m4}. Recently, it has been showed that in this expression  the relative positions and clock times of the detectors before and after the radiation transit differ by a BMS supertranslation \cite{strom2,strom4}; indeed, the displacement memory formula is shown to be equivalent to Weinberg's
formula for soft graviton production. 

In this section we study the phenomenon of gravitational memory in an expanding universe and show that, as in the flat Minkowski background,  it can still be connected to a BMS transformation.
 Gravitational memory, which  provides an alternative way of proving soft theorems in flat spacetime,   might also be useful in finding   consistent relations for cosmological correlators \cite{kr}, and this  is the main motivation for studying this phenomenon here.


Free falling objects follow spacetime geodesics and the spatial separation  between two nearby geodesics will  change with time in general as a result of  the spacetime curvature. Let us consider then two objects sitting in two  nearby points in a small region perturbed by a gravitational wave burst. 
The curvature tensor for the gravitational wave is oscillating and it is expected that the separation of the two nearby points oscillates as well. Indeed, the separation vector $X^i$ between two points will be wiggling as the wave train is passing by.  
As we are interested in the gravitational memory  which is a property of the asymptotic gravitational field at null infinity, it is enough to consider the deviation caused to nearby geodesics due to the Weyl tensor.   The reason for this is that we we will use a perturbative approach to the problem and in first order perturbation,  the Weyl tensor provides a gauge invariant way to do this. Therefore, for free falling objects at nearly points with spatial separation $X^i$, we will have 
 \begin{eqnarray}
\frac{d^2X^i}{\d\tau^2}=-C^i_{0j0} X^j \left(\frac{\d X^0}{\d\tau}\right)^2,\label{gdw}
\end{eqnarray}
where $C^\mu_{\nu\rho\sigma}$ is the Weyl tensor \cite{BG,BGY,chu,Bieri1}. 
Then, Eq. (\ref{gdw}) is explicitly written as 
  \begin{eqnarray}
\frac{\d^2X^i}{\d \tau^2}=\frac{1}{2}\partial^2_{\tau\tau} {h^i}_{j}^{\rm TT} X^j, \label{gdw1}
\end{eqnarray}
where $h^{\rm TT}_{ij}$ stands for the transverse traceless part of the metric perturbation.
 By integrating Eq. (\ref{gdw1}),  we get that the induced shift  in  the  displacement of two nearby points 
 \begin{eqnarray}
 \label{po}
 \Delta X^i=\frac{1}{2}X^j  \Delta {h^i}_{j}^{\rm TT},
 \end{eqnarray}
where we have denoted by $\Delta X^i $ and $\Delta h_{ij}^{\rm TT}$  the shift in $X^i$ and 
$ h_{ij}^{\rm TT}$  due to  the gravitation wave train.

\subsection{The gravitational memory: standard approach}
We present first the standard way of computing the  change $\Delta h_{ij}^{\rm TT} $.
Let us assume that the four-momentum of a mass 
$m$ moving along a trajectory $s^\mu(\lambda)$ in the rest frame of a distant observer at ${\cal O}$ is $p^\alpha=m\,u^\alpha$ and $k^\alpha$ is the past-directed null four-vector from observer to the source. 
\begin{figure}[h!]
\begin{center}
\includegraphics[scale=.6]{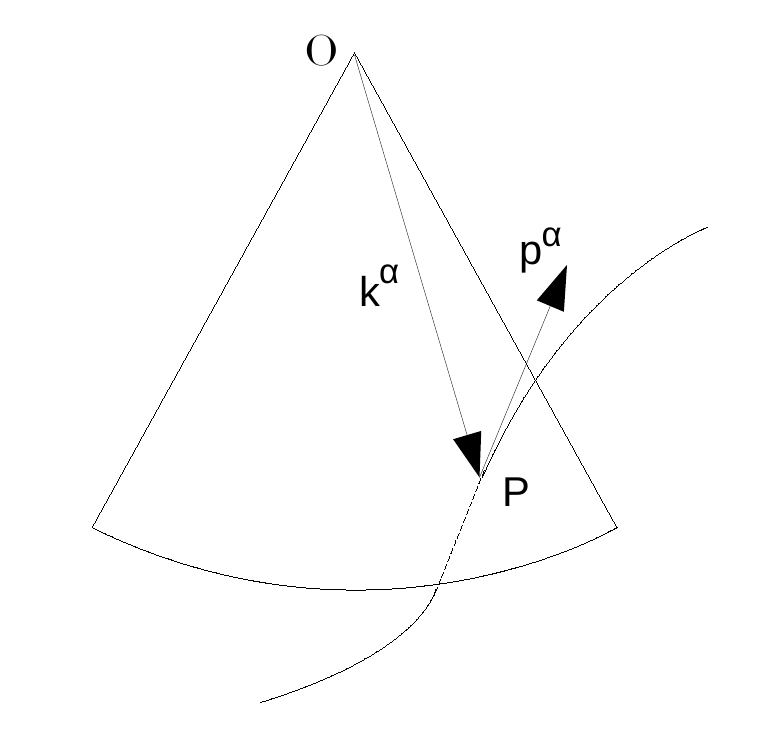}
\caption{\small Geometry of light cone.}
\end{center}
\end{figure}
Then the motion of the source is described by the action 
\begin{eqnarray}
 {\mathcal S}=-m \int \d\lambda \sqrt{-g_{\mu\nu}(s(\lambda))u^\mu(\lambda) u^\nu(\lambda)}
 \end{eqnarray} 
and the resulting energy-momentum tensor will be given by 
\begin{eqnarray}
T^{\mu\nu}=\frac{m}{\sqrt{-g_{\mu\nu}(x)}}\int \d\lambda \frac{u^\mu(\lambda)u^\nu(\lambda)\delta^{(4)}\left(x-s(\lambda)\right)}{\sqrt{-g_{\mu\nu}(s(\lambda))u^\mu(\lambda) u^\nu(\lambda)}}. \label{emtp}
\end{eqnarray}
Since $\delta {T^{i\,{\rm TT}}_j} =0$ for perfect fluids, the only source for graviton perturbation will be 
the moving particle and so we will have 
\begin{eqnarray}
{h^{\rm TT}_{ij}}''+2{\cal H} {h_{ij}^{\rm TT}}'-\nabla^2h^{\rm TT}_{ij}=16\pi G a^2  {T}_{ij} ^{\rm TT}. \label{tt}
\end{eqnarray}
The solution of Eq. (\ref{tt})  can be expressed as 
\begin{eqnarray}
h_{ij}^{\rm TT}=-\sqrt{4\pi G}\left[\int \d^4x' a^4(\tau')\, G^R_{ij\rho\sigma}(x,x') T^{\rho\sigma}(x')\right]^{\rm TT}.   \label{hTT}
\end{eqnarray}
For example, the energy momentum tensor of  $n$ freely moving point particles with velocities $u^\mu$ which collide at $\tau=0$ and then move apart with velocities ${u'}^\mu$  is 
\begin{eqnarray}
T^{\mu\nu}=\frac{1}{\sqrt{-g_{\mu\nu}(x)}}\int \d\lambda\, \delta^{(4)}\left(x-s(\lambda)\right)
\left(\sum_{A=1}^nm_A \frac{u_A^\mu(\lambda)u_A^\nu(\lambda)}{\sqrt{-g_{\mu\nu}u_A^\mu u_A^\nu}}~\theta(-\tau)+ m_A\frac{{u'}_A^\mu(\lambda){u'}_A^\nu(\lambda)}{\sqrt{-g_{\mu\nu}{u'}_A^\mu {u'}_A^\nu}}~\theta(\tau)\right). \label{emtpp}
\end{eqnarray}
In this case, Eq. (\ref{hTT}) implies
\begin{eqnarray}
h_{ij}^{\rm TT}=-m\sqrt{4\pi G}\left[\int \d\lambda \, G^{\rm R}_{ij\rho\sigma}\big{(}x,s(\lambda)\big{)}\sum_{A=1}^n
m_A u_A^\rho\big{(}s(\lambda)\big{)}u_A^\sigma\big{(}s(\lambda)\big{)}\varepsilon_A\right]^{\rm TT}. \label{hTT1}
\end{eqnarray}
where $\varepsilon_A=\pm$ for incoming and outgoing particles, respectively. 
The kernel $G^{\rm R}_{\mu\nu\rho\sigma}$ above is just the retarded Green function. It is  derived  in Appendix C and for a matter dominated FRW universe is explicitly written as

\begingroup
\allowdisplaybreaks
\begin{eqnarray}
G^{\rm R}_{\mu\nu\rho\sigma}(x,x')&=&
\overline G^{\rm R}_{\mu\nu\rho\sigma}(x,x')+\widehat G^{\rm R}_{\mu\nu\rho\sigma}(x,x')\nonumber \\
&=&
\frac{\theta(\Delta\tau)}{2\pi a(\tau)a(\tau')}\delta\Big{(}(x-x')^2\Big{)}\Big{(} \eta_{\mu\rho}\eta_{\nu\sigma}+\eta_{\mu\sigma}\eta_{\nu\rho}-\eta_{\mu\nu}\eta_{\rho\sigma} \Big{)}\nonumber\\
&-&
\frac{\theta(\Delta\tau)}{4\pi a(\tau)a(\tau')}\frac{1}{\tau\tau'}\theta(|\Delta\tau|-\Delta x)\left(P^{(1)}_{\mu\nu\rho\sigma}+\frac{3}{5}P^{(3)}_{\mu\nu\rho\sigma}\right)\nonumber \\
&+&\frac{3\theta(\Delta\tau)}{8\pi a(\tau)a(\tau')}\frac{1}{\tau^2\tau'^2}\left[\frac{3}{2}(x-x')^2-\frac{3}{4}\tau\tau'\right]\theta(|\Delta\tau|-\Delta x) P^{(2)}_{\mu\nu\rho\sigma}\label{green-ret} 
\nonumber\\
&-&\frac{\theta(|\Delta\tau|-\Delta x)}{20\pi a(\tau)a(\tau')}\frac{\theta(\Delta \tau)}{\tau^3\tau'^3}\left[\frac{15}{4}(x-x')^4-5 (x-x')^2 \tau
\tau'+12 \tau^2 \tau'^2\right]\ P^{(3)}_{\mu\nu\rho\sigma},\nonumber\\
&&
\end{eqnarray}
\endgroup
where $\overline G^{\rm R}$ is the part of the retarded Green function supported on the light cone (proportional to the  $\delta$-function) and $\widehat G^{\rm R}$ is the rest. 
We observe that the retarded Green function is not zero inside  the light cone. Indeed, the first line in Eq. (\ref{green-ret}) is the usual form of the retarded Green function in Minkowski spacetime (up to the scale factors). It represents signals traveling strictly on the light cone at a speed of light and determines the field at some point $x^\mu$ 
specified by the source at some retarded time $\Delta \tau=\Delta \vec{x}$.  
However,  we see that there are  additional terms in the Green function which are all proportional to $\theta(|\Delta \tau|-\Delta \vec{x})$. This means that there are also signals traveling in the interior of the light cone. All these signals represent what is known as ``tail". This is a violation of Huygens principle similarly to what   happens to  wave propagation in odd spacetime
 dimensions. In even spacetime dimensions, a light pulse  passing through a point  lasts for a short time and then fades out immediately. The light wavefront has both a sharp
 front and a sharp rear. In odd spacetime dimensions, the light  pulse  reaches the point at some time,  but it does not die out immediately, on the contrary it does  have a tail. Waves in odd spacetime dimensions have a sharp front, but diffused rear. 
This  difference between the two cases is mathematically  explained by recalling that, on the complex plane, the wave moves on $\Delta\vec{x}^2 +(it-z)^2=0$. This has singularities at  $z=it\pm \Delta\vec{x}$. In even spacetime dimensions, this singularity is a pole, whereas in odd spacetime dimensions it is a branch cut. 

This behavior is present also in a  FRW background. The existence of a tail for waves in curved spacetimes is known and it has been noted in Ref. \cite{dW}. In particular, even the dynamics of a single charge  depends on its  past history and,  in addition to the force present in Minkowski spacetime, there is another force which is exerted on the charge upon itself. This force is due to the waves which are emitted away from the charge  and then scatter back by the geometry and interact with the charge at some later time. This is the reason why  in general waves that travel in curved backgrounds are not on the light-cone, but have a tail that fills the interior of the light cone as well.

 Since the retarded propagator  can be split into two parts, one representing  propagation on the light cone and one propagation in its interior, the tensor mode  will also be characterized by two pieces,   the standard pulse and  the 
tail contribution.
For example in the case of $n$-colliding free moving point particles we considered above, each particle will follow a geodesic with  four-velocity $u^\mu$ given in conformal coordinates by
\begin{eqnarray}
u^\mu=\left(\frac{\d \tau}{\d \lambda}=\gamma,\frac{\d x_i}{\d\lambda}=\frac{v_i}{a^2}\right), ~~~\gamma(\lambda)=\frac{\sqrt{a^2-v_i^2}}{a^2} \label{uu}
\end{eqnarray}
and  integral curve  $s^a(\lambda)=\Big{(}\tau(\lambda),x^i(\lambda)\Big{)}$, 
\begin{eqnarray}
 \tau(\lambda)=\int^\lambda \gamma(\lambda')\d \lambda', ~~~x^i(\lambda)=\frac{v^i \tau(\lambda)}{qa }+x^i_0.  \label{uuu}
 \end{eqnarray} 
Then, we may split the tensor mode as
\begin{eqnarray}
h_{ij}^{\rm TT}=\overline h _{ij}^{\rm TT}+\widehat h _{ij}^{\rm TT},
\end{eqnarray}
where 
$\overline  h _{ij}^{TT}$ is the part originating from the propagation on the light-cone 
\begin{eqnarray}
\overline h _{ij}^{\rm TT}=
-\sqrt{4\pi G}\left[\int \d^4x' a^4(\tau')\, \overline G^{\rm R}_{ij\rho\sigma}(x,x') T^{\rho\sigma}(x')\right]^{\rm TT},   \label{hTTO}
\end{eqnarray}
and $\widehat h_{ij}^{\rm TT}$ is the part originating from the tail  
\begin{eqnarray}
\widehat h_{ij}^{\rm TT}=-\sqrt{4\pi G}\left[\int \d^4x' a^4(\tau')\, \widehat G^{\rm R}_{ij\rho\sigma}(x,x') T^{\rho\sigma}(x')\right]^{\rm TT}.   \label{hT}
\end{eqnarray}
Using the energy-momentum given in  Eq. (\ref{emtpp}) and  Eqs. (\ref{uu}) and (\ref{uuu}), we find that 
\begin{eqnarray}
\overline  h _{ij}^{\rm TT}=-\left(\frac{G}{4\pi}\right)^{1/2}\frac{1}{a(\tau)}\sum_{A=1}^n\left(\frac{p^A_i p^A_j|_{\lambda_0}}{\big{(}x^a-s^a(\lambda_0)\big{)} p^A_\alpha}\varepsilon_A\right)^{\rm TT}, 
\end{eqnarray}
where $p^A_i=m_A u_i^A$ is the momentum of the particles and $\lambda_0$ is the solution of the equation
\begin{eqnarray}
\Big{(}x-s(\lambda_0)\Big{)}^2=0.
\end{eqnarray}

On the other hand, for $\widehat h_{ij}^{\rm TT}$ we find that 
\begin{eqnarray}
\widehat   h _{ij}^{\rm TT}=\left(\frac{G}{4\pi}\right)^{1/2}\frac{1}{a(\tau)\tau }
\left(\sum_{A=1}^n m_A v^A_i v_j^A\right)^{\rm TT}  \left(
\int_{\tau_{\rm m}}^{\tau_0}
\d\tau'\frac{\gamma(\tau')}{\tau'a^5(\tau')}
+\cdots \right) \label{h00}
\end{eqnarray}
where $\tau_{\rm m}$ is the time when matter domination of the universe started,  $\tau_0$ is 
the solution of the equation
\begin{eqnarray}
 \tau-\tau_0-|x^i-s^i(\tau_0)|=0, \label{co}
 \end{eqnarray} 
and  the dots stands for the rest of the contributions.  Clearly, 
 for radial motion, $\tau_0$ should solve 
 \begin{eqnarray}
  \tau-r-\tau_0+|s^i(\tau_0)|=0,
  \end{eqnarray} 
 and therefore $\tau_0=\tau_0(u)$ is a function of the retarded coordinate $u=\tau-r$. Then Eq. (\ref{h00}) is written as 
\begin{eqnarray}
\widehat   h _{ij}^{\rm TT}=\left(\frac{G}{4\pi}\right)^{1/2}\frac{1}{a(\tau)\tau }
\left(\sum_{A=1}^n m_A v^A_i v_j^A\right)^{\rm TT}  \bigg({\rm I}(u)
+\cdots \bigg), \label{h000}
\end{eqnarray}  
 where 
 \begin{eqnarray}
 {\rm I}(u)=\int_{\tau_{\rm m}}^{\tau_0(u)}
\d\tau'\frac{\gamma(\tau')}{\tau'a^5(\tau')}.
 \end{eqnarray}
 We observe that at null infinity the part of the tensor perturbations due to the tail of the retarded Green function gets an extra ${\cal O}(1/r)$ suppression compared to the light-cone propagation and hence to leading order we will have  
 \begin{eqnarray}
 h _{ij}^{\rm TT}= 
 -\left(\frac{G}{4\pi}\right)^{1/2}\sum_{A=1}^n\left(\frac{p^A_i p^A_j}{\big{(}\overline x^a-\overline r^a(\lambda_0)\big{)} p^A_\alpha}\varepsilon_A\right)^{\rm TT},
 \end{eqnarray}
 where $\overline x^a-\overline s^a(\lambda_0)=a(\tau)\Big{(}x^a-s^a(\lambda_0)\Big{)}$ is the physical distance from the source. 
 Therefore, the gravitational memory in a cosmological background is the same as in flat spacetime at null infinity (to leading order) up to the redshift factors. We will confirm this  in the next subsection  by a direct calculation of the memory using the BMS symmetry. 

\subsection{The gravitational memory from BMS}
Let us now discuss the gravitational memory phenomenon in terms of Bondi coordinates for observers at null infinity.
We start from the  length of a vector $X^\mu$ connecting two geodesics changes. We know that the geodesic deviation is written as 
\begin{eqnarray}
\frac{D^2X^\mu}{\d\lambda^2}=-R^\mu_{\nu\rho\sigma}X^\rho u^\nu u^\sigma,
\end{eqnarray}
where $u^\mu$ is the tangent to the geodesic and as usual
\begin{eqnarray}
 \frac{D}{\d\lambda}=u^\mu \nabla_\mu.
 \end{eqnarray} 
Then for the length $X^2=X^\mu X_\mu$ we will have 
\begin{eqnarray}
\frac{D^2X^2}{\d\lambda^2}=2g_{\mu\nu}X^\mu \frac{D^2X^\nu}{\d\lambda^2}+2 g_{\mu\nu}
\frac{DX^\mu }{\d\lambda}\frac{DX^\nu}{\d\lambda} = 2 X \frac{\d^2 X}{\d\lambda^2}+2 \left(\frac{\d X}{\d\lambda}\right)^2,
\end{eqnarray}
from which we find that 
\begin{eqnarray}
X \frac{\d^2 X}{\d\lambda^2}=-R_{\mu\rho\nu\sigma}X^\mu X^\nu u^\rho u^\sigma +\Xi, \label{dis}
\end{eqnarray}
where 
\begin{eqnarray}
\Xi= g_{\mu\nu}
\frac{D X^\mu }{\d\lambda}\frac{D X^\nu}{\d\lambda}- \left(\frac{\d X}{\d\lambda}\right)^2=
g_{\mu\nu}
\frac{D X^\mu }{\d\lambda}\frac{D X^\nu}{\d\lambda}-\frac{1}{X^2}X_\mu X_\nu \frac{D X^\mu }{\d\lambda}\frac{D X^\nu}{\d\lambda}.
\end{eqnarray}
Let us consider now BMS detectors traveling at fixed radius $r_0$ and fixed angle ($z=z_0,\bz=\bz_0$) along the trajectory 
\begin{eqnarray}
X^\mu_{\rm BMS}(u)=\left(u,r_0,z_0,\bz_0\right). 
\end{eqnarray}
It is easy then to check that the vector 
\begin{eqnarray}
V^\mu_{\rm BMS}=\frac{\d X^\mu_{\rm BMS}}{\d\lambda}=\left(\frac{\d u}{\d\lambda},0,0,0\right)
\end{eqnarray}
satisfies the following relation
\begin{eqnarray}
V^\mu_{\rm BMS}\nabla_\mu V^\nu_{\rm BMS}=a_0^{-1}\left(\frac{q}{r_0},0,0,0\right), ~~~V_{\rm BMS}^\mu \,g_{\mu\nu}V_{\rm BMS}^\nu =-1
\end{eqnarray}
for
\begin{eqnarray}
\frac{\d u}{\d \lambda}=\frac{1}{a_0} ,~~~ a_0=(r_0/L)^q.
\end{eqnarray}
Thus, to leading order in $1/r_0$, $X^\mu_{\rm BMS}$ is an inertial trajectory. 
Let us now find the change in the length of a shift vector $\xi^\mu$ for two nearby BMS detectors sitting at the same $r=r_0$ and at $z_1=0,z_2=\delta z$. We may assume that the detectors are along a meridian so that $X^\mu=(0,0,\delta z,\delta \bz)$ with $\delta z=\delta\bz$. Notice that if they are not aligned  along a meridian, we may rotate  $\delta z$ and $\delta \bz$ according to $\delta z\to \delta ze^{-i\phi}$ where $e^{2i\phi}=\delta \bz/\delta z$, and correspondingly $C_{zz}$ is rotated as 
\begin{eqnarray}
C_{zz}\to C_{zz}e^{-2i\phi}=C_{zz}\frac{\delta z}{\delta \bz}.  \label{rotC}
\end{eqnarray}
It is easy to see that in this case we have  $\Xi=0$ and Eq. (\ref{dis}) reduces to  
\begin{eqnarray}
 \frac{\d^2 X}{\d\lambda^2}=-\omega\,  X, \label{dis1}
\end{eqnarray}
where 
\begin{eqnarray}
\omega=R_{\mu\rho\nu\sigma}\hat X^\mu\hat X^\nu u^\rho u^\sigma
\end{eqnarray}
and  $\hat X^\mu=X^\mu/X$ is the unit displacement vector. 
By using 
\begin{eqnarray}
u^\mu=V_{\rm BMS}^\mu, ~~~\hat X^{z}=1/\sqrt{g_{z\bz}}
\end{eqnarray}
and  the relevant non-zero component of the Riemann tensor in  leading order in $r_0$ 
\begin{eqnarray}
R_{uzuz}=-\frac{1}{2}a_0^2r_0\partial_u^2C_{zz},
\end{eqnarray}
we find  
\begin{eqnarray}
\omega=-\frac{(1+z\bz)^2}{8r_0a_0^2}\left(\partial_u^2C_{zz}+{\rm c.c.}\right).
\end{eqnarray}
Eq. (\ref{dis1}) is then   explicitly written as 
\begin{eqnarray}
 \frac{\d^2X}{\d u^2}=
 \frac{(1+z\bz)^2}{8r_0}\left(\partial_u^2C_{zz}+{\rm c.c}\right)
  \, X. 
\end{eqnarray}
It is solved by 
\begin{eqnarray}
X(\lambda)=X(0)\bigg[1+\frac{(1+z\bz)^2}{8r_0}\left(C_{zz}+C_{\bz\bz}\right)
\bigg],  \label{DX0}
\end{eqnarray}
where $X(0)$ is the initial length of the displacement vector.
Therefore, the gravitational wave has  pulled apart the BMS detectors by an amount 
$\Delta X=X(\lambda_2)-X(\lambda_1)$ which turns out to be
\begin{eqnarray}
\Delta X=(1+z\bz)^2\frac{X(0)}{8r_0}\left(\Delta C_{zz}
\frac{\delta z}{\delta \bz}
+\Delta C_{\bz\bz}
\frac{\delta \bz}{\delta z}\right). \label{DX}
\end{eqnarray}
 after using Eq. (\ref{rotC}) for  arbitrary  $\delta z$ and $\delta \bz$. This equation corresponds to Eq. (\ref{po}) where  $C_{zz}$
represents the traceless and transverse part of the metric.

 Eq. (\ref{DX}) for the memory can alternatively be calculated by performing a BMS transformation \cite{strom3,strom4} which produces the following shift 
 
\be
(z,\bz)\to (z+\delta z,\bz+\delta \bz)= \left(z+\frac{D^zf}{r},\bz +\frac{D^\bz f}{r}\right).
\ee
If $C_{zz}=0$, this shift  produces a $\Delta C_{zz}=2 D_z^2 f$ as discussed in Appendic C. 
Correspondingly, the length $X^2=g_{z\bz}\delta z \delta \bz$ gets shifted as
 
 \begin{eqnarray}
\Delta X= \xi^\mu\del_\mu X=\frac{1}{2X}\xi^\mu\del_\mu X^2. \label{DXBMS}
\end{eqnarray}
A straightforward  calculation of Eq. (\ref{DXBMS}) reproduces  Eq.(\ref{DX}), as expected. In order to explicitly find the memory $\Delta X$, one needs to solve for $\Delta C_{zz}$. For instance, if one takes  $N=C_{zz}=0$,  $\Delta C_{zz}$ is calculated along the lines  of  Ref. \cite{strom3}. By using Eqs. (\ref{DC}) and (\ref{Tuu}) in Appendix A, we explicitly find 
\begin{eqnarray}
\Delta C_{zz}(z,\bz)=\frac{4}{\pi}\int \d^2\zeta \gamma_{\zeta\overline \zeta}\frac{\bz-\overline \zeta}{z-\zeta},
\frac{(1+\zeta\overline z)^2}{(1+\zeta \bz)(1+z\bz)^3}  A(\zeta,\bar \zeta),
\end{eqnarray}
where 
\begin{eqnarray}
A(\zeta,\overline \zeta)=2(1+q)(1+2q)\Delta m+(1+2q)8\pi G \int_{u_i}^{u_f}\d u \, 
 \delta T_{uu}^{(2)}. 
\end{eqnarray}
This equation reduces to the flat case when $q=0$. 
\section{Conclusions}
Symmetries provide  a powerful guiding principle in cosmology. The de Sitter isometry group has been employed to characterize the properties of  the  cosmological perturbations  generated by an  inflationary stage. For instance,  the Goldstone boson of a spontaneously broken dilation symmetry is associated to  the scalar fluctuation in single-field inflation.  Symmetry arguments allow also to write consistency relations among the  large-scale structure correlator functions. They appear under the form of 
 soft-pion theorems relating 
an $n$-point function to an $(n+1)$-point function where the additional leg represents the emission (or absorption) of the 
 Goldstone boson associated with a non-linearly realized symmetry. 

In observational cosmology light rays emitted by distant objects are the carriers of the information about the statistical distributions
of the large-scale structure. This means that photons from cosmological sources reach observers at null infinity.  One might wonder if the BMS symmetry play any role in cosmology by reproducing these signals through a BMS transformation, leading to a corresponding
soft theorem or Ward identity. 

Motivated by such arguments, in this paper we have discussed 
the BMS transformations in a cosmological setting. 
 In particular, we have  studied  the structure of asymptotic past and future infinity in cosmological FRW  spacetimes. Our investigation  is limited to decelerating FRW backgrounds where future null infinity exists.  We have identified the BMS transformations acting   at future null infinity and leaving the asymptotic metric invariant to leading order.  We have found that the corresponding  cosmological Gosldstone modes  and identified them in  scalar, vector  and tensor degrees of freedom as those degrees of freedom which may exist at  null infinity and perturb the asymptotic data. Hence cosmological BMS transformations generate physically inequivalent vacua as the null infinity  is populated with physical degrees of freedom.

As gravitational memory and BMS symmetry are intimately connected \cite{strom3,strom4}, 
we have also discussed  the phenomenon of gravitational memory when cosmological expansion is taken into account.  We have found that  there are extra contributions to the gravitational memory due to the tail of the retarded Green functions of the FRW background. Indeed the retarded Green functions  are supported not only on the light-cone, but also in its interior. We will discuss the implications of our findings in a forthcoming publication.

\vskip.3in 

\noindent
 \section*{Acknowledgments}
 We would like to thank Dieter L\"ust for discussions. 
A.R. is supported by the Swiss National
Science Foundation (SNSF), project `Investigating the Nature of Dark Matter" (project number: 200020${}_{-}$159223).

\begin{appendices}

\renewcommand{\theequation}{A.\arabic{equation}}
 \setcounter{equation}{0}
 
 \section{Einstein equations}
Einstein equations can  be written near null infinity perturbatively in the $1/r$ expansion. We  assume  that the form of the energy-momentum tensor is
\begin{eqnarray}
T_{\mu\nu}=\overline{T}_{\mu\nu}+\delta T_{\mu\nu}, \label{exp1}
\end{eqnarray}
where $\overline{T}_{\mu\nu}$ is the matter energy-momentum tensor  supporting  the FRW background and $\delta T_{\mu\nu}$ is the perturbation (including gravity self-sourcing).  
It is straightforward to find that  the non-zero components of $\overline{T}_{\mu\nu}$ for the FRW metric  in Eq. (\ref{dseta11}) are
\begin{eqnarray}
\overline{T}_{uu}&=&\overline{T}_{ur}=\frac{1}{8\pi G} \frac{3q^2}{(r+u)^2},\nonumber \\
\overline{T}_{rr}&=&\frac{1}{8\pi G} \frac{2q(1+q)}{(r+u)^2},\nonumber\\
\overline{T}_{z\bz}&=&
\frac{1}{8\pi G}\frac{2-q}{(r+u)^2} r^2 \gamma_{z\bz}. 
\end{eqnarray}
At null infinity, these components turns out to be
\begin{eqnarray}
\overline{T}_{uu}&=&\overline{T}_{ur}=\frac{1}{8\pi G} \frac{3q^2}{r^2}, \nonumber\\
\overline{T}_{rr}&=&\frac{1}{8\pi G} \frac{2q(1+q)}{r^2},
\nonumber \\
\overline{T}_{z\bz}&=&
\frac{1}{8\pi G}(2-q)\gamma_{z\bz}\bigg(1-\frac{2u}{r}+\frac{3u^2}{r^2}\bigg).
\end{eqnarray}

We will adopt an expansion of $\delta T_{\mu\nu}$ of the form
\begin{eqnarray}
\delta T_{\mu\nu}=\delta T_{\mu\nu}^{(0)}+\frac{\delta T^{(1)}_{\mu\nu}}{r}+\frac{\delta T^{(2)}_{\mu\nu}}{r^2}+\cdots. \label{exp4}
\end{eqnarray}
Then, for the metric Eq. (\ref{dseta1111}), we find that Einstein equations are written as
\begingroup\allowdisplaybreaks
\begin{eqnarray}
-(1+q)\del_uN-\del_u^2C_{z\bz}+2\partial_u E&=&8\pi G \delta T^{(1)}_{uu},\nonumber\\
-2 (1+q)\del_u m-\del_u\Big{(}D^z U_z+D^\bz U_\bz\Big{)}
-(1+4q)N-D^2N+(1+2q)\del_uC_{z\bz}\nonumber\\
+2\partial_u F+2D^2 E-2(1+2q)E
&=&  8\pi G \delta T_{uu}^{(2)}\label{Tuu},\nonumber\\
-(1+q)D_z E&=&8\pi G\delta T^{(1)}_{rz},\nonumber\\
(1+4q) N+(1+2q) \del_u C_{z\bz}+D^2E-2E-q(2+q)E
&=&8\pi G \delta T^{(2)}_{ur}, \nonumber \\
-q \del_u N +\del_u U_z+\frac{1}{2}\del_u D^z C_z-\frac{1}{2}\del_u D_z C_{z\bz}&=& 8 \pi G \delta T^{(1)}_{uz}, \nonumber\\
\frac{1}{2}D_z C_{z\bz}-(1+2q)U_z-\frac{1}{2}D^z C_{zz}-\frac{1}{2}(3+2q) D_z F +q u E&=& 8\pi G  \delta T^{(2)}_{rz},\nonumber\\
qN_{zz}&=&8\pi G\delta T^{(0)}_{zz},\nonumber\\
q(q-2) C_{zz}+2 q D_z U_z-q u \del_u C_{zz}+D_z^2 F&=&8\pi G\delta T^{(1)}_{zz},\nonumber\\
\left(-q^2 N-q \del_u C_{zz}\right)/\gamma^{z\bz}&=&8\pi G \delta T^{(0)}_{z\bz},\nonumber\\
\gamma_{z\bz}\Big{(}q (2-q) C_{z\bz}+2q(2-q)m\Big{)}-q(D_\bz U_z+D_z U_\bz)-D_zD_\bz F-\gamma_{z\bz}(F+2 q u E)&=&8\pi G \delta T^{(1)}_{z\bz}.\nonumber\\
&&
\end{eqnarray}
\endgroup
Let us note that under the BMS transformation (\ref{BMS}), the energy-momentum  transform as 
$T_{\mu\nu}\to T_{\mu\nu}+\delta  \widetilde T_{\mu\nu}$ , where

\begin{eqnarray}
\delta  \widetilde T_{\mu\nu}={\cal L}_\xi T_{\mu\nu}.
\end{eqnarray}
Therefore, by using Eqs. (\ref{exp1}--\ref{exp4}) and expanding $\delta  \widetilde T_{\mu\nu}$ in powers of $1/r$ as in Eq. (\ref{exp4}), we get for example

\begin{eqnarray}
\delta \widetilde T_{uu}^{(1)}&=&-f \del _u \delta  T_{uu}^{(1)}+D_z \delta  T_{uu}^{(0)}D^z f+
D_\bz \delta T_{uu}^{(0)}D^\bz f, \nonumber \\
\delta \widetilde T_{uu}^{(2)}&=&\delta  T_{uu}^{(1)}D^zD_zf+D_z \delta T_{uu}^{(1)}D^z f+
D_\bz \delta  T_{uu}^{(1)}D^\bz f, \nonumber \\
\delta \widetilde T_{ur}^{(2)}&=& \delta T_{ur}^{(1)}D^zD_zf+D_z \delta  T_{ur}^{(1)}D^z f+
D_\bz \delta T_{uu}^{(1)}D^\bz f-\delta  T_{uz}^{(0)}D^z f-\delta T_{u\bz}^{(0)}D^\bz f, \nonumber \\
\delta \widetilde T_{uz}^{(1)}&=&-f \del_u \delta T_{uz}^{(1)}-\delta  T_{uu}^{(1)}D_z f+\delta T_{uz}^{(0)}D_zD^zf f-\delta T_{u\bz}^{(0)}D_\bz D^\bz f\nonumber \\
&-&\delta  T_{ur}^{(1)}D^zD_zf+D_z \delta T_{uz}^{(0)}D^z f+
D_\bz \delta  T_{uz}^{(0)}D^\bz f
,\nonumber \\
\delta \widetilde T_{rz}^{(2)}&=&
2 \delta T_{rz}^{(1)} D_z D^z f-\delta T_{zz}^{(0)}D^z f-\delta  T_{z\bz}^{(0)}D^\bz f+\delta T_{r\bz}^{(1)}D^\bz D_z f\nonumber \\
&+&D_z \delta  T_{rz}^{(1)}D^z f+D_\bz \delta T_{rz}^{(1)} D^\bz f
-2q \frac{(1+q)}{8\pi G}\bigg(D^zD_z^2f+D_z f\bigg),
\nonumber \\
\delta \widetilde T_{zz}^{(0)}&=&
-\delta  T_{uz}^{(0)}D_z f-\delta  T_{rz}^{(0)}D_zD^zD_zf-f \del_u \delta  T_{zz}^{(0)},
\nonumber \\
\delta \widetilde T_{zz}^{(1)}&=&-f \del_u\delta  T_{zz}^{(1)}+2 \delta T_{zz}^{(0)}D_z D^z f+2 \delta T_{z\bz}^{(0)}D^\bz D_z f-2 \delta  T_{uz}^{(1)}D_z f\nonumber \\
&+&D_z \delta  T_{zz}^{(0)}D^z f
+D_\bz \delta  T_{zz}^{(0)}D^\bz  f-2 D_z \delta  T_{rz}^{(1)}D_zD^z D_z f
+2\frac{(2-q)q}{8\pi G}D_z^2 f.
\end{eqnarray}


\section{Retarded Green functions}
\renewcommand{\theequation}{B.\arabic{equation}}
 \setcounter{equation}{0}
We may calculate the graviton propagator from the gravitational action after gauge fixing. 
For this, we consider the action 
\begin{eqnarray}
{\cal S}=\int \d^4 x \sqrt{-g}\left(\frac{1}{16\pi G} R+{\cal L}_{\rm matter}\right),
\end{eqnarray}
where  ${\cal L}_{\rm matter}$ is the matter energy-momentum tensor of a perfect fluid. We may expand around the FRW background    as 
\begin{eqnarray}
 g_{\mu\nu}=a^2(\tau)\left(\eta _{\mu\nu}+\sqrt{16\pi G}\, h_{\mu\nu}\right)
 \end{eqnarray} 
and by adding the gauge fixing term 
\begin{eqnarray}
{\cal L}_{\rm GF}=-\frac{a^2(\tau)}{2}\left(\partial_\mu h^\mu_\nu-\frac{1}{2}\partial_\mu h^\nu_\nu
-2\frac{a'}{a} h_{0\mu }\right)^2 
\end{eqnarray}
we find that the quadratic part for the metric fluctuations is written as 
\begin{eqnarray}
{\cal L}_{2}=\frac{1}{2}h^{\mu\nu} {D_{\mu\nu}}^{\rho\sigma}h_{\rho\sigma}.
\end{eqnarray}
We define then the graviton propagator as \cite{A,AT,TS}
\begin{eqnarray}
i\Delta^q_{\mu\nu\rho\sigma}(x,x')=\Big< 0\left|T\Big{(}h_{\mu\nu}(x)h_{\rho\sigma}(x')\Big{)}\right|0\Big>.
\end{eqnarray}
The propagator is written in terms of the tensors 
\begin{eqnarray}
P^{(1)}_{\mu\nu\rho\sigma}&=&2\left(P_{\mu(\rho}P_{\sigma)\nu}-P_{\mu\nu}P_{\rho\sigma}\right),\\
P^{(2)}_{\mu\nu\rho\sigma}&=&-4 n_{(\mu}P_{\nu)(\rho}n_{\sigma)},\\
P^{(3)}_{\mu\nu\rho\sigma}&=&\left(P_{\mu\nu}+n_{\mu}n _\nu\right)
\left(P_{\rho\sigma}+n_{\rho}n_\sigma\right),
\end{eqnarray}
where 
\begin{eqnarray}
P_{\mu\nu}=\eta_{\mu\nu}+n_\mu n_\nu, ~~~n^\mu n^\nu\eta_{\mu\nu}=-1.
\end{eqnarray}
Explicitly we have 
\begin{eqnarray}
i\Delta^q_{\mu\nu\rho\sigma}(x,x')=i\Delta^q_1(x,x')\left(P^{(1)}_{\mu\nu\rho\sigma}+\frac{1+q}{1+2q}P^{(3)}_{\mu\nu\rho\sigma}\right)+i\Delta^q_2(x,x') P^{(2)}_{\mu\nu\rho\sigma}
+i\Delta^q_3(x,x')q P^{(3)}_{\mu\nu\rho\sigma},\nonumber\\
&&
\end{eqnarray}
with  $i\Delta^q_i$ given by 
\begin{eqnarray}
i\Delta^q_i(x,x')&=&\int \frac{\d^3k}{(2\pi)^3}e^{i\vec{k}\cdot (\vec{x}-\vec{x}')}
\left[\theta(\tau-\tau')\Psi^{(i)}_k(\tau)\overline \Psi ^{(i)}_k(\tau')+
\theta(\tau'-\tau)\overline \Psi^{(i)}_k(\tau)\Psi ^{(i)}_k(\tau)\right]\nonumber \\
&=& \frac{1}{2\pi^2\Delta x}\int_0^\infty \d k\, k \, \sin (k \Delta x)
\left[\theta(\Delta\tau)\Psi^{(i)}_k(\tau)\overline \Psi ^{(i)}_k(\tau')+
\theta(-\Delta\tau)\overline \Psi^{(i)}_k(\tau)\Psi ^{(i)}_k(\tau)\right],\nonumber \\
&&
\end{eqnarray}
with  $\Delta \tau=\tau-\tau', ~ \Delta x=|\vec{x}-\vec{x}'|$ 
and $\Psi ^{(i)}_k(\tau)$ satisfy the following equations. 
\begingroup
\allowdisplaybreaks
\begin{eqnarray}
&&\left(\partial_\tau^2+\frac{2q}{\tau}\partial_\tau+k^2\right)\Psi ^{(1)}_k(\tau)=0,\\
&&\left(\partial_\tau^2+\frac{2q}{\tau}\partial_\tau-\frac{2q}{\tau^2}+k^2\right)\Psi ^{(2)}_k(\tau)=0,\\
&&\left(\partial_\tau^2+\frac{2q}{\tau}\partial_\tau-\frac{1+2q}{\tau^2}+k^2\right)\Psi ^{(3)}_k(\tau)=0.
\end{eqnarray}
\endgroup
The retarded Green function $G^{\rm R}_{\mu\nu\rho\sigma}(x,x')$ are  given by 
\begin{eqnarray}
G^{\rm R}_{\mu\nu\rho\sigma}(x,x')=2\theta(\Delta \tau){\rm Im}\biggl[i\Delta^q_{\mu\nu\rho\sigma}(x,x')\biggr]
\end{eqnarray}
and therefore we will have 
\begin{eqnarray}
G^{\rm R}_{\mu\nu\rho\sigma}(x,x')&=&G^q_1(x,x')\left(P^{(1)}_{\mu\nu\rho\sigma}+\frac{1+q}{1+2q}P^{(3)}_{\mu\nu\rho\sigma}\right)\nonumber\\
&+&G^q_2(x,x') P^{(2)}_{\mu\nu\rho\sigma}
+G^q_3(x,x') \frac{q}{1+2q} P^{(3)}_{\mu\nu\rho\sigma},
\end{eqnarray}
where 
\begin{eqnarray}
G^q_i(x,x')=2\theta(\Delta \tau){\rm Im}\biggl[i\Delta^q_i(x,x')\biggr].
\end{eqnarray}
The explicit form of the retarded Green function depends on the background. For example, for matter dominance where $q=2$ we will have 

\begingroup
\allowdisplaybreaks
\begin{eqnarray}
{\rm Im}\biggl[i\Delta^{(2)}_1(x,x')\biggr]&=&\frac{1}{4\pi a(\tau)a(\tau')}\delta\Big{(}(x-x')^2\Big{)}-\frac{1}{8\pi a(\tau)a(\tau')}\frac{1}{\tau\tau'}\theta(|\Delta\tau|-\Delta x),\\
{\rm Im}\biggl[i\Delta^{(2)}_2(x,x')\biggr]&=&\frac{1}{4\pi a(\tau)a(\tau')}\delta\Big{(}(x-x')^2\Big{)}\nonumber\\
&&+\frac{3}{16\pi a(\tau)a(\tau')}\frac{1}{\tau^2\tau'^2}\left\{\frac{3}{2}(x-x')^2-\frac{3}{4}\tau\tau'\right\}\theta(|\Delta\tau|-\Delta x), \\
{\rm Im}\biggl[i\Delta^{(2)}_3(x,x')\biggr]&=&\frac{1}{4\pi a(\tau)a(\tau')}\delta\Big{(}(x-x')^2\Big{)}\nonumber
\\
&-&\frac{\theta(|\Delta\tau|-\Delta x)}{16\pi a(\tau)a(\tau')}\frac{1}{\tau^3\tau'^3}\left\{\frac{15}{4}(x-x')^4-5 (x-x')^2 \tau
\tau'+12 \tau^2 \tau'^2\right\}.\nonumber\\
&&
\end{eqnarray}
\endgroup
The retarded Green function is   written as \cite{TS}
\begingroup
\allowdisplaybreaks
\begin{eqnarray}
G^{\rm R}_{\mu\nu\rho\sigma}(x,x')&=&2\theta(\Delta\tau)\left(P^{(1)}_{\mu\nu\rho\sigma}+\frac{3}{5}P^{(3)}_{\mu\nu\rho\sigma}\right)\left\{\frac{1}{4\pi a(\tau)a(\tau')}\delta\Big{(}(x-x')^2\Big{)}\right.\nonumber \\&&
\left. -\frac{1}{8\pi a(\tau)a(\tau')}\frac{1}{\tau\tau'}\theta(|\Delta\tau|-\Delta x)\right\}+2\theta(\Delta\tau) P^{(2)}_{\mu\nu\rho\sigma}\nonumber \\
&&\times \left\{\frac{3}{16\pi a(\tau)a(\tau')}\frac{1}{\tau^2\tau'^2}\left[\frac{3}{2}(x-x')^2-\frac{3}{4}\tau\tau'\right]\theta(|\Delta\tau|-\Delta x)\right. \nonumber \\
&&\left.+\frac{1}{4\pi a(\tau)a(\tau')}\delta\Big{(}(x-x')^2\Big{)}\right\}
+\frac{4}{5}\theta(\Delta \tau)P^{(3)}_{\mu\nu\rho\sigma}\nonumber\\
&&\times \left\{-\frac{1}{16\pi a(\tau)a(\tau')}\frac{1}{\tau^3\tau'^3}\left[\frac{15}{4}(x-x')^4-5 (x-x')^2 \tau
\tau'+12 \tau^2 \tau'^2\right]\theta(|\Delta\tau|-\Delta x)\right.\nonumber\\
&&\left.+\frac{1}{4\pi a(\tau)a(\tau')}\delta\Big{(}(x-x')^2\Big{)}\right\}. \label{ggr}
\end{eqnarray} 
\endgroup
Using the identity
\begin{eqnarray}
P^{(1)}_{\mu\nu\rho\sigma}+P^{(2)}_{\mu\nu\rho\sigma}+P^{(3)}_{\mu\nu\rho\sigma}=
\eta_{\mu\rho}\eta_{\nu\sigma}+\eta_{\mu\sigma}\eta_{\nu\rho}-\eta_{\mu\nu}\eta_{\rho\sigma},
\end{eqnarray}
 the retarded  Green function in Eq. (\ref{ggr})  is written as in Eq. (\ref{green-ret}). 

\section{The Newman-Penrose formalism}
\renewcommand{\theequation}{C.\arabic{equation}}
 \setcounter{equation}{0}
We will present here the NP construction which is particularly appropriate for the discussion of gravitational radiation.  
In the  NP formalism we may define null tetrads $\ell^\mu,n^\mu,m^\mu,\overline m ^\mu$  at leading order as
\begin{eqnarray}
\ell^\mu&=&\left(\frac{r+u}{L}\right)^{-2q}\bigg(0,1,0,0\bigg),\label{np}\\
n^\mu&=&\bigg(
1,-\frac{1}{2}+\frac{N}{2}+\frac{m}{r},\frac{U^z}{r},\frac{U^\bz}{r}
\bigg),\nonumber\\
m^\mu&=&\left(\frac{r+u}{L}\right)^{-q}\frac{1+z\bz}{\sqrt{2}\, r}\left(0,0,
1-\frac{C_{z\bz}}{r},-\frac{{C^\bz}_z}{r}
\right). \nonumber
\end{eqnarray}
It is straightforward to verify that these null tetrads satisfy the relations
\begin{eqnarray}
&&\ell_\mu \ell^\mu=n_\mu n^\mu=m_\mu m^\mu=0,\ell_\mu m^\mu=\ell_\mu m^\mu=n_\mu m^\mu=n_\mu\overline m^\mu=0, \nonumber \\
&&\ell_\mu n^\mu=-1,~~~m_\mu \overline m ^\mu=1, ~~~g^{\mu\nu}=-\ell^\mu n^\nu-\ell^\nu n^\mu +
m^\mu \overline m ^\mu +m^\nu \overline m ^\mu. 
\end{eqnarray}
By employing the null tetrads, we can define the Weyl-NP scalars 
\begin{eqnarray}
\Psi_0&=&C_{\mu\nu\rho\sigma}\ell^\mu m^\nu \ell^\rho m ^\sigma, \nonumber \\
\Psi_1&=&C_{\mu\nu\rho\sigma}\ell^\mu n^\nu \ell^\rho m ^\sigma, \nonumber \\
\Psi_2&=&C_{\mu\nu\rho\sigma}\ell^\mu m^\nu \overline m ^\rho n ^\sigma, \nonumber \\
\Psi_3&=&C_{\mu\nu\rho\sigma}\ell^\mu n^\nu \overline m ^\rho n ^\sigma, \nonumber \\
\Psi_4&=&C_{\mu\nu\rho\sigma}n^\mu \overline m^\nu n^\rho\overline m ^\sigma. \label{psi}
\end{eqnarray}
In particular, we find that to leading order in $r$
\begin{eqnarray}
\Psi_4&=&-\frac{1}{2}\frac{\del_u {N_z}^z}{r}+{\cal O}(r^{-2}),\\ 
\Psi_2&=&\frac{1}{a^2}\frac{\Psi_2^{(2)}}{r^2}+\frac{1}{a^2}\frac{\Psi_2^{(3)}}{r^3}+{\cal O}(r^{-4}), 
\label{psi24}
\end{eqnarray}
where 
\begin{eqnarray}
\Psi_2^{(2)}&=&-\frac{1}{6}\left(N+\del_u C_{z\bz}\right),\nonumber\\
\Psi_2^{(3)}&=&-m+\frac{1}{4}C^{zz}N_{zz}-\frac{2}{3}D^zU_z+\frac{1}{3}D^\bz U_\bz
-\frac{1}{12}D^zD^zC_{zz}-\frac{1}{12}D\bz D^\bz C_{\bz\bz}\nonumber \\
&&+\frac{1}{6}D^zD_z C_{z\bz}
+\frac{qu}{3}\left(N+\del_u C_{z\bz}\right).
\end{eqnarray}
We may impose appropriate conditions at $\sc^+_\pm$ which read
\begin{eqnarray}
N_{zz}|_{\sc^+_\pm}=0.
\end{eqnarray}
In particular, we will have for the 
 imaginary part of $\Psi_2$ 
\begin{eqnarray}
{\rm Im} \Psi_2|{\sc^+_+}|_{\sc^+_\pm}={\rm Im} \Psi_2^{(3)}|_{\sc^+_\pm}=\left(D^zU_z-D^\bz U_\bz\right)|_{\sc^+_\pm}=0,
\end{eqnarray}
which is written as 
\begin{eqnarray}
\left(D_\bz^2C_{zz}-D_z^2C_{\bz\bz}\right)|_{\sc^+_\pm}=0.
\end{eqnarray}
Therefore, in the vacuum, where $C_{zz}$ is $u$-independent
we will have that  \cite{strom1,strom2}
\begin{eqnarray}
D_\bz^2C_{zz}-D_z^2C_{\bz\bz}=0,
\end{eqnarray}
which is solved for 
\begin{eqnarray}
C_{zz}=-2D_z^2 C(z,\bz). \label{DC}
\end{eqnarray}

\end{appendices}

\end{document}